\documentclass[aps,pra,twocolumn,groupedaddress,floatfix]{revtex4}

\usepackage{amsmath}
\usepackage{graphicx}
\usepackage{multirow}
\usepackage{color}

\begin{document}

\title{Measuring nonequilibrium retarded spin-spin Green's functions in an ion-trap based quantum simulator}
\author{Bryce T. Yoshimura and J. K. Freericks}
\affiliation{Department of Physics, Georgetown University, 37th and O st. NW, Washington DC, 20057, USA}

\date{\today}

\begin{abstract}

Recent work proposed a variant on Ramsey interferometry for coupled spin-$1/2$ systems that directly measures the retarded spin-spin Green's function. We expand on that work by investigating nonequilibrium retarded spin-spin Green's functions within the transverse-field Ising model. We derive the lowest four spectral moments to understand the short-time behavior and we employ a Lehmann-like representation to determine the spectral behavior. We simulate a Ramsey protocol for a nonequilibrium quantum spin system that consists of a coherent superposition of the ground state and diabatically excited higher-energy states via a temporally ramped transverse magnetic field. We then apply the Ramsey spectroscopy protocol to the final Hamiltonian, which has a constant transverse field. The short-time behavior directly relates to Lieb-Robinson bounds for the transport of many-body correlations, while the long-time behavior relates to the excitation spectra of the Hamiltonian. Compressive sensing is employed in the data analysis to efficiently extract that spectra.

\end{abstract}

\pacs{to be determined}
\date{\today}

\maketitle

\section{Introduction} 

One of the main computational tools of quantum many-body physics are the retarded Green's functions because their causal structure makes them the physical Green's functions for the linear response of the system and for describing its equilibrium behavior. The formal interpretation of the retarded Green's functions is that they determine the quantum states of the system. In equilibrium, all single-particle expectation values can be calculated from the Green's functions and the Fermi-Dirac distribution, which determines how those quantum states are occupied. If one is also interested in exciting systems to nonequilibrium, then one needs to examine nonequilibrium many-body theory. Here, one has to determine two independent Green's functions---the retarded Green's function (introduced above) and the so-called lesser Green's function---the former continuing to determine the quantum states and the latter determining how those states are occupied (since it is no longer given by a simple Fermi-Dirac distribution). It turns out that nearly all experimentally measurable quantities are actually determined by the lesser Green's function, not the retarded Green's function. This holds in equilibrium too, as one typically finds any expectation value calculated from the retarded Green's function requires an additional Fermi-Dirac distribution factor, which converts the retarded Green's function into the lesser Green's function. It is due to this simple relationship between the retarded and lesser Green's functions in equilibrium that one can get all the information from knowing the retarded Green's functions only.

\begin{figure}
	\centering
	\begin{tabular}{ c }
		\includegraphics[scale=0.33]{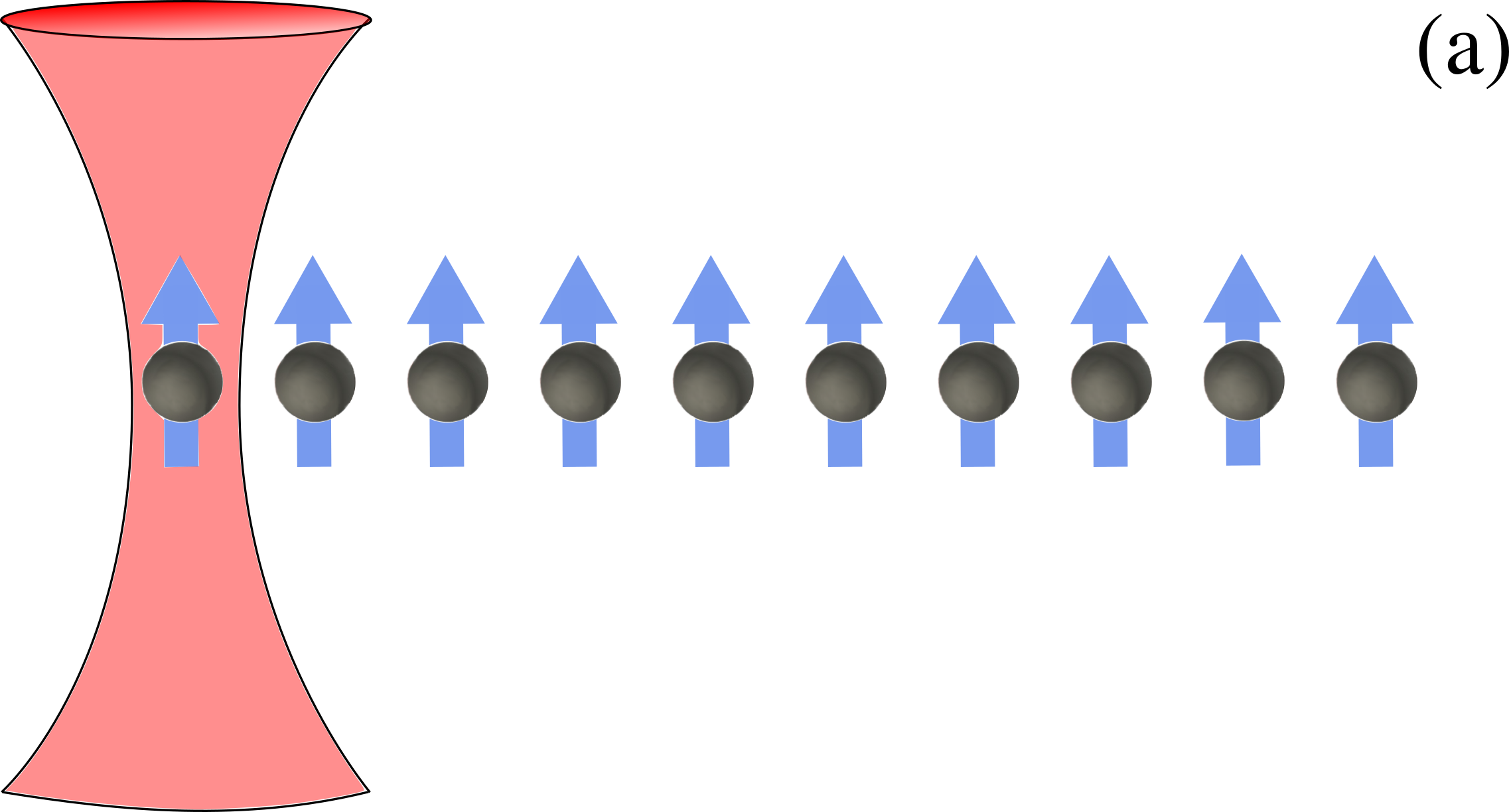}\\
		\includegraphics[scale=0.33]{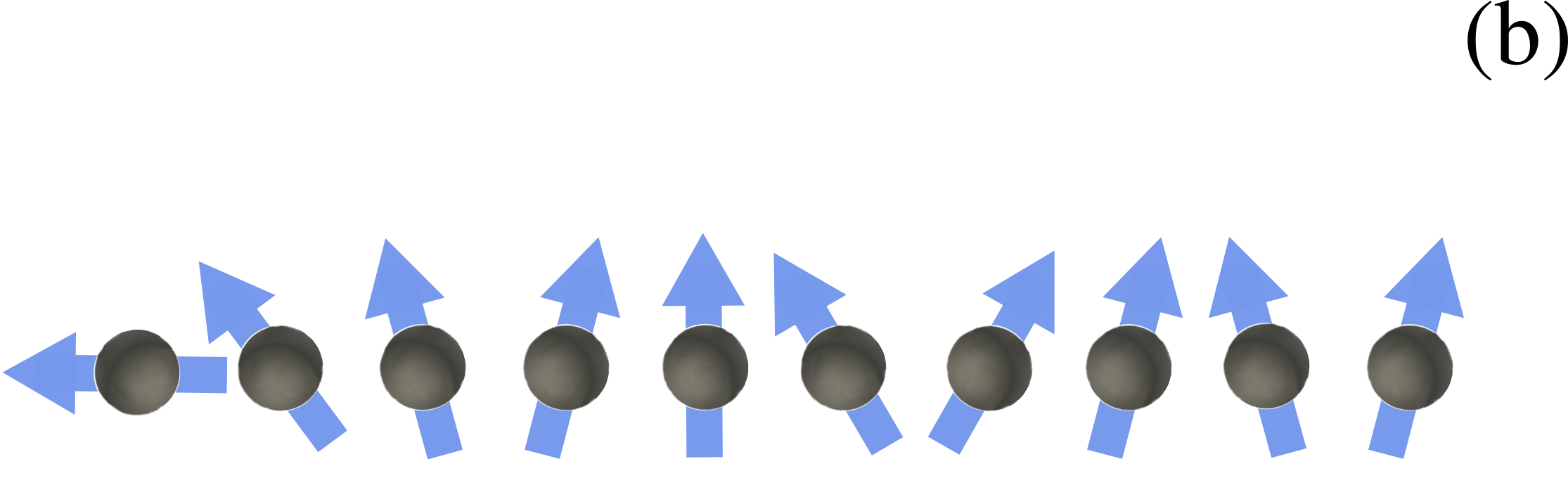}\\
\\
\\
		\includegraphics[scale=0.08]{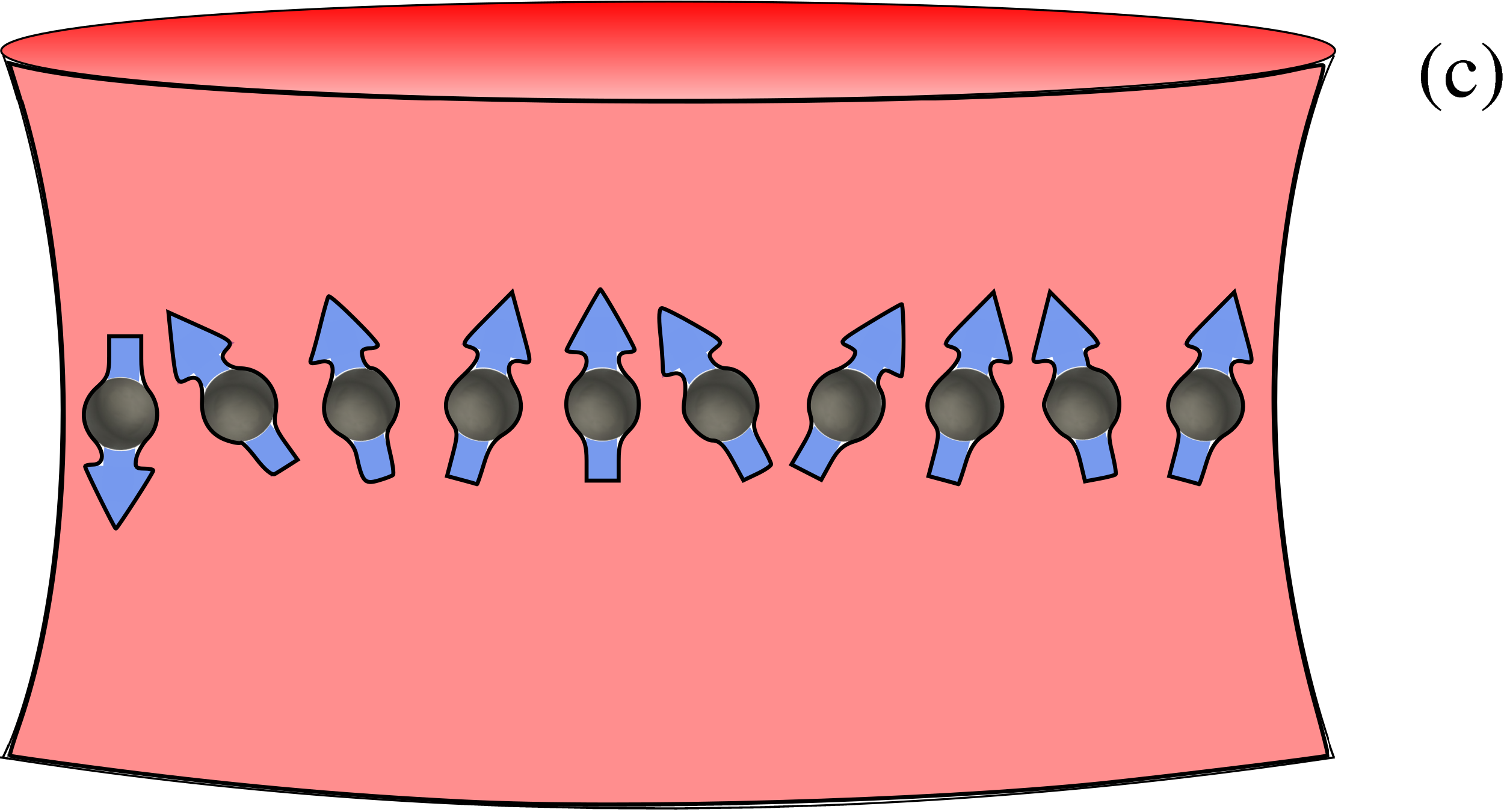}
	\end{tabular}
	\caption{(Color online.) Schematic of the Ramsey protocol. (a) Rotate a single spin $j$ by $\pi/2$. (b) Allow the resulting quantum spin state to freely evolve forward in time. (c) Apply a global rotation and immediately measure the $z$-component of the $i^{\text{th}}$ spin, $\sigma^{(z)}_i$. }
	\label{fig:Ramsey_Protocol}

\end{figure}

This then brings up a fundamental question: Is it possible to directly measure the retarded Green's function in an experiment? Most physicists would reasonably respond no, since the occupancy of the states
always plays a role in a measurement, but recent work showed that this is not the case. Knap, {\it et al.}~\cite{Knap}, proposed a variant of Ramsey interferometry for any coupled spin-$1/2$ system, that reduces to the {\it direct experimental measurement of a retarded Green's function}! The Ramsey protocol is quite simple, as shown in Fig.~\ref{fig:Ramsey_Protocol}. One starts the system in some given quantum state (Knap {\it et al.} assume this is a thermal state, but in most quantum simulators it will be some other state that the system has evolved into at time $t=t_0$), applies a local Rabi pulse at site $j$, lets the system evolve under the Hamiltonian until time $t$, applies a second  global Rabi pulse, and then measures the spin at site $i$. It is by no means obvious that this will result in the measurement of the retarded Green's function, so we illustrate this in a brief derivation below. Then we describe how the evolution of the system will change (from that of an equilibrium Green's function) due to the nonequilibrium character of the initial state $|\psi_0\rangle$ that the system started from. We will also describe how this method can be employed to examine Lieb-Robinson~\cite{lieb_robinson} bounds. 

The organization of this paper is as follows: In Sec.~II, we summarize the derivation of how the Ramsey spectroscopy protocol results in a pure state retarded spin-spin Green's function. We review the formalism of the transverse-field Ising model as simulated in the linear Paul trap and the approximation we apply for the time evolution. To further understand the short time behavior the spectral moments are derived. We finally discuss the key ideas of compressive sensing. In Sec.~III, we present numerical examples of the pure state retarded spin-spin Green's function. We extract different features as a function of time to examine Lieb-Robinson bounds and we apply compressive sensing to Fourier transform the measurements as a function of time and extract the excitation spectra at different transverse fields. Finally, in Sec.~IV, we provide our conclusions.

\section{Formalism}

The Ramsey spectroscopy protocol is completely general, so we first describe it solely in terms of spins, and then we discuss the specific implementation via ions trapped in a linear Paul trap (for the concrete calculations).
The procedure involves applying two rotations of the spins ({\it i.~e.}, Rabi pulses) at different times, with a free evolution under the spin Hamiltonian in between; the first rotation is a single-spin rotation at lattice site $j$, given by
\begin{equation}
	R_j(\phi_1 ) = \frac{1}{ \sqrt{2} } \left[ \hat{I} + i \left(\sigma^{(x)}_j \cos\phi_1 - \sigma^{(y)}_j \sin\phi_1 \right) \right].
\end{equation} 
Here, $\sigma^{(r)}_j$ is the Pauli spin matrix (with eigenvalues $\pm1$), the index $r = x$, $y$, or $z$  denotes the spatial direction of the Pauli spin matrix and the index $j$ denotes the spatial site index on the lattice.  The second spin rotation is a global spin rotation given by
\begin{equation}
	R( \phi_2 ) = \prod^N_{j=1} R_j( \phi_2 )
\end{equation}
for a spin lattice with $N$ lattice sites. The Rabi pulse is the general one used in Ref.~\onlinecite{Knap},
with the product of the Rabi frequency times the time equal to $\pi/2$, and $\phi_1$ (or $\phi_2$) the phase of the laser pulse.

The Pauli matrices satisfy the standard commutation relations
\begin{equation}
	\left[\sigma^{(\alpha)}_i, \sigma^{(\beta)}_j \right]_- = 2i \epsilon_{\alpha \beta \gamma} \sigma^{(\gamma)}_i \delta_{ij}
	\label{eq:spincommutation}
\end{equation}
with $\epsilon$ the completely antisymmetric rank three tensor (Levi-Civita symbol).

 The pure state retarded spin-spin Green's function is defined by
\begin{equation}
	G_{\alpha \beta , ij}^{\rm ret}(t,t_0) = -i\theta(t-t_0) \langle \psi_0 | \left[\sigma^{(\alpha)}_i(t), \sigma^{(\beta)}_j(t_0) \right]_- | \psi_0 \rangle,
	\label{eq:Greens} 
\end{equation}
where $\theta(t)$ is the Heaviside function, and $|\psi_0\rangle$ is a pure quantum state which can be thought of as the ``initial'' spin wavefunction. This is a ``nonequilibrium'' Green's function, similar to the $T=0$ Green's function, except it uses a different quantum state than the ground-state for the matrix elements; for example, in an ion-trap-based implementation, it can be the time-evolved state when the system starts in the ground state for a large magnetic field and then the field is ramped to some final value. Note that the time evolution between the two spin rotations can be with respect to a constant Hamiltonian or a time-varying one, it does not matter for the definition.  In addition, the initial state $|\psi_0\rangle$ is taken to be any pure quantum state; it need not be an eigenstate of the Hamiltonian at the initial time.

\subsection{ Ramsey spectroscopy protocol}

The Ramsey spectroscopy protocol consists of four steps after starting the system in an initial state $|\psi_0\rangle$ at time $t_0$: (1) perform a single-spin rotation on the $j^\text{th}$ spin, with the single-spin rotation $R_j(\phi_1)$ at $t_0$, (2) evolve the system to time $t$ under the Hamiltonian (which can be time dependent, but will be chosen to be time-independent here), (3) perform a global rotation $R(\phi_2)$ at time $t$, and (4) immediately measure the $z$-component of the $i^\text{th}$ spin. The entire Ramsey interferometry measurement then corresponds to evaluating the following matrix element ($t\ge t_0$)
\begin{equation}
	M_{i,j}( \phi_1, \phi_2, t) = \langle \psi(t) | \sigma^{(z)}_i | \psi(t) \rangle,
\end{equation}
where $|\psi(t)\rangle$ is the Schr\"odinger representation for the final wavefunction (after the first three steps of the protocol), which is given by 
\begin{equation}
	| \psi(t ) \rangle = R(\phi_2) \hat{U}( t ,t_0) R_j( \phi_1) | \psi_0\rangle. 
	\label{eq:stateprep}
\end{equation} 
Here, $\hat{U}(t,t_0)=\mathcal{T}_t \exp[-i\int_{t_0}^t d\bar t \mathcal{H}(\bar t)]$ is the evolution operator, given by a time-ordered product if the Hamiltonian changes as a function of time. Using the fact that
\begin{equation}
R^\dagger(\phi_2)\sigma_i^{(z)}R(\phi_2)=-\sigma_i^{(x)}\sin\phi_2-\sigma_i^{(y)}\cos\phi_2,
\end{equation}
and the Heisenberg representation for the spin operators, where $\sigma_i^{(r)}(t)=\hat{U}^\dagger(t,t_0)\sigma_i^{(r)}\hat{U}(t,t_0)$, yields
\begin{eqnarray}
M_{i,j}( \phi_1, \phi_2, t) &=&-\frac{1}{2} \langle \psi_0|[\hat{I}-i(\sigma_j^{(x)}\cos\phi_1-\sigma_j^{(y)}\sin\phi_1)]\nonumber\\
&\times&[\sigma_i^{(x)}(t)\sin\phi_2+\sigma_i^{(y)}(t)\cos\phi_2]\nonumber\\
&\times&
[\hat{I}+i(\sigma_j^{(x)}\cos\phi_1-\sigma_j^{(y)}\sin\phi_1)]|\psi_0\rangle.
\end{eqnarray}
The most interesting case corresponds to the choice $\phi_1=0$ and $\phi_2 = \pi/2$~\cite{Knap} which gives
\begin{eqnarray}
		M_{i,j}( 0, \frac{\pi}{2},t ) &=& -\frac{1}{2}\langle \psi_0 | [\hat{I}-i\sigma_j^{(x)}(t_0)]\sigma_i^{(x)}(t)\nonumber\\
&\times&[\hat{I}+i\sigma_j^{(x)}(t_0)]|\psi_0\rangle,
\end{eqnarray}
where we trivially represented the spins in $R_j(\phi_1)$ by the Heisenberg representation at $t_0$, since $\hat U(t_0,t_0)=\hat I$.
So far, the Ramsey protocol, and the manipulations we have made, are completely general. Now, we need to invoke a parity argument that says all expectation values that correspond to an odd number
of $\sigma_i^x$ operators vanish if the initial state $|\psi_0\rangle$ has definite spin-reflection parity. We will be considering the evolution with respect to a transverse-field Ising model, which has this spin-reflection parity (and will be verified in detail below). In this case, the matrix element becomes
\begin{equation}
M_{i,j}( 0, \frac{\pi}{2},t )=\frac{1}{2}G_{xx,ij}^{\rm ret}(t,t_0)
\label{eq:ramseysimp}
\end{equation}
after we drop the odd averages.
Hence, the Ramsey spectroscopy directly measures the retarded spin-spin Green's function! Note that this also implies that the Green's function in the time domain is real (which can be easily proven).

A natural alternative representation of the retarded Green's function is the Lehmann representation and here we derive a similar representation when the transverse-field Ising model is time independent during the free-evolution stage of the Ramsey spectroscopy. We first expand $|\psi_0\rangle = \sum_n C_n | n\rangle$, in terms of the eigenstates of the transverse-field Ising model  at time $t_0$ ($\mathcal{H}(t_0)|n\rangle = E_n | n\rangle$); the Hamiltonian becomes time independent for $t\ge t_0$. We introduce $|\psi_0\rangle$ and an identity operator into Eq.~(\ref{eq:Greens}),
\begin{equation}
\begin{aligned}
	G_{xx,ij}^{\rm ret}&(t,t_0) =  -i\theta(t-t_0) \\
	& \times \sum^N_{m,n, n'} C^*_m C_n \left[ \langle m | \sigma^{(x)}_i(t)|n'\rangle\langle n'| \sigma^{(x)}_j(t_0)| n \rangle \right.\\
	 & -\left. \langle m|\sigma^{(x)}_j(t_0)|n'\rangle\langle n'| \sigma^{(x)}_i(t) | n \rangle\right].
\end{aligned}
\end{equation}     
Because we are assuming that the Hamiltonian is time independent, the time evolution operator acting on an eigenstate satisfies $\hat U(t,t_0)| n\rangle = \exp[ -i E_n (t - t_0) ]| n\rangle$, which is employed to further simplify the above equation to
\begin{equation}
\begin{aligned}
	&G_{xx,ij}^{\rm ret}(t,t_0) =  -i\theta(t-t_0) \\
	& \times \sum_{m,n,n'} C^*_m C_n \left[  \text{e}^{-i( E_{n'} - E_m) ( t -t_0) } \langle m | \sigma^{(x)}_i|n'\rangle\langle n'| \sigma^{(x)}_j| n \rangle \right.\\
	 & -\left. \text{e}^{ -i( E_n - E_{n'}) ( t -t_0) } \langle m|\sigma^{(x)}_j|n'\rangle\langle n'| \sigma^{(x)}_i | n \rangle\right].
\end{aligned}
\label{eq:lehmann}
\end{equation}  
In this representation, the individual matrix elements oscillate at the energy differences of the transverse-field Ising model. Although some matrix elements might cancel each other once summed over, the pure state retarded Green's function will oscillate at many energy eigenvalue differences. Additionally the energy differences are between states with opposite spin-reflection parity, because the $\sigma^{(x)}$ operator is odd under the spin reflection symmetry. Interestingly, this equation differs from the conventional Lehmann representation of a thermal Green's function because the matrix element is not proportional to $|\sigma^{(\alpha)}|^2$. In other words, when we evaluate the pure state Green's function---in cases where the state is a superposition of eigenstates---the Lehmann representation includes cross terms that do not appear in the conventional trace (when one evaluates a thermally averaged Green's function).

\subsection{Transverse-field Ising model}

For concreteness, we consider the evolution of the spin system in the transverse-field Ising model as generated in an ion trap quantum simulator. In the linear Paul trap, the effective spin-$1/2$ system is encoded onto the $^2S_{1/2}:|F=0, m_f=0\rangle$ and $|F=1, m_f=0\rangle$ hyperfine ``clock" states of the trapped $^{171}$Yb$^+$ ion. The Ising-like interaction is generated by applying two optical beams with a frequency difference of $\mu$, which results in a spin-dependent force. When the phonons are only virtually occupied, they can be integrated out to leave behind a spin-only Hamiltonian.  These Ising spins  have a long-range interaction that decays approximately with a power law in the inter-ion distance. The power law is tunable between the uniform case ($\alpha=0$) and the dipole-dipole interaction case ($\alpha=3$). The spin-exchange interactions are approximated by
\begin{equation}
|J_{ij}| \approx \frac{ J_0 }{ |R_i - R_j |^{\alpha} }
\label{eq:approx}
\end{equation} 
where $R_i$ is the position of the $i^\text{th}$ ion and $J_0$ the overall scale for the exchange interactions. The $J_{ij}$s are positive for the ferromagnetic case and negative for the antiferromagnetic case; we will show results only for the ferromagnetic case here. In general, the $J_{ij}$s are time dependent but when the detuning $\mu$ is detuned to the blue of the transverse center-of-mass mode, the system is well approximated by static $J_{ij}$, which we do here as well.

The transverse-field Ising model for $N$ ions then becomes
\begin{equation}
\mathcal{H}(t) = - \sum_{\scriptsize\begin{array}{c}
i,j=1\\
i<j
\end{array}}^N J_{ij} \sigma^{(x)}_i \sigma^{(x)}_j - B^{(y)}(t) \sum_{i = 1}^N \sigma^{(y)}_i.
\end{equation}
Here, $B^{(y)}(t)$ is the time-dependent transverse magnetic field; we work with $\hbar = 1$. Note, that the transverse-field Ising model has two symmetries. The first symmetry is a spatial reflection symmetry, which is described by $J_{ij} = J_{N-iN-j}$, and derives from the even symmetry of the axial trapping potential  about the origin. Here, the lattice index is ordered in a strictly increasing order from left to right. The second symmetry is a spin-reflection parity, when the Pauli matrices are transformed  by $\sigma^{(x)} \to -\sigma^{(x)}$, $\sigma^{(y)} \to \sigma^{(y)}$, $\sigma^{(z)} \to -\sigma^{(z)}$. In this case, the spin-spin commutators and the transverse-field Ising model Hamiltonian both remain invariant. So, if the initial state $|\psi_0\rangle$ has a definite spin-reflection parity, then, because the Hamiltonian $\mathcal{H}(t)$ is
even under spin-reflection parity (and hence so is the evolution operator), we have that the matrix element with respect to $|\psi_0\rangle$ of any odd number of $\sigma_i^{(x)}(t)$ operators vanishes, as claimed above. 

\begin{figure}

	\includegraphics[scale=0.08]{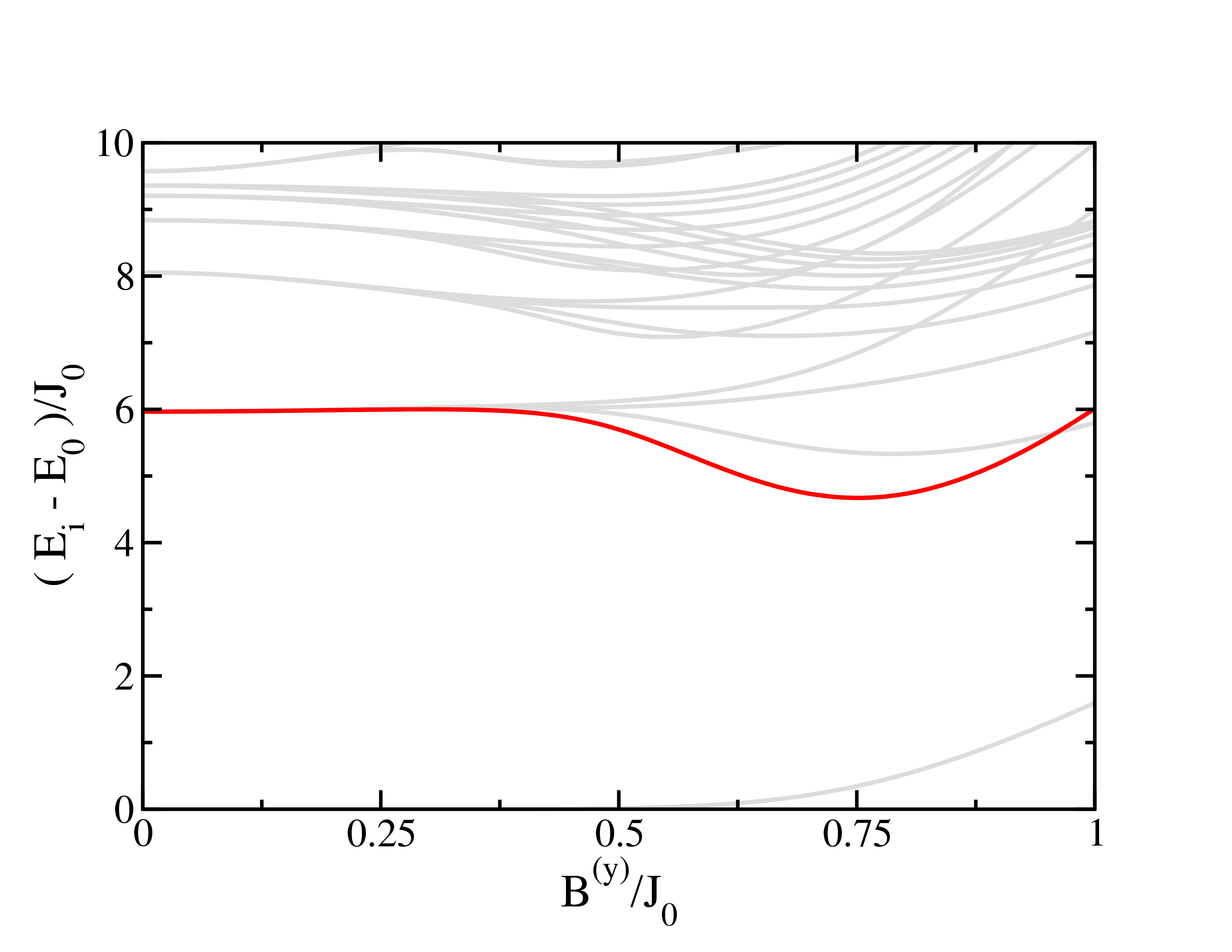}
	
	\caption{(Color online.) Energy spectrum of the transverse-field Ising model with $10$ spins and $\alpha = 1.00$. Near $B^{(y)}/J_0 = 0.75$ there is a minimum energy gap between the ground state and the (red line) first coupled state. }
	\label{fig:energy}

\end{figure}

The excitations to higher coupled states ({\it i.~e.}, states with the same symmetry) depend on the energy gaps between those coupled states and how quickly the transverse magnetic field changes with time up to the time $t_0$. In Fig.~\ref{fig:energy}, we show the energy spectra for $N=10$ spins with $\alpha=1.00$; note that there is a minimum energy gap near $B^{(y)}/J_0 = 0.75$.

\begin{figure}
	\centering
	\begin{tabular}{ c }
		\includegraphics[scale=0.35]{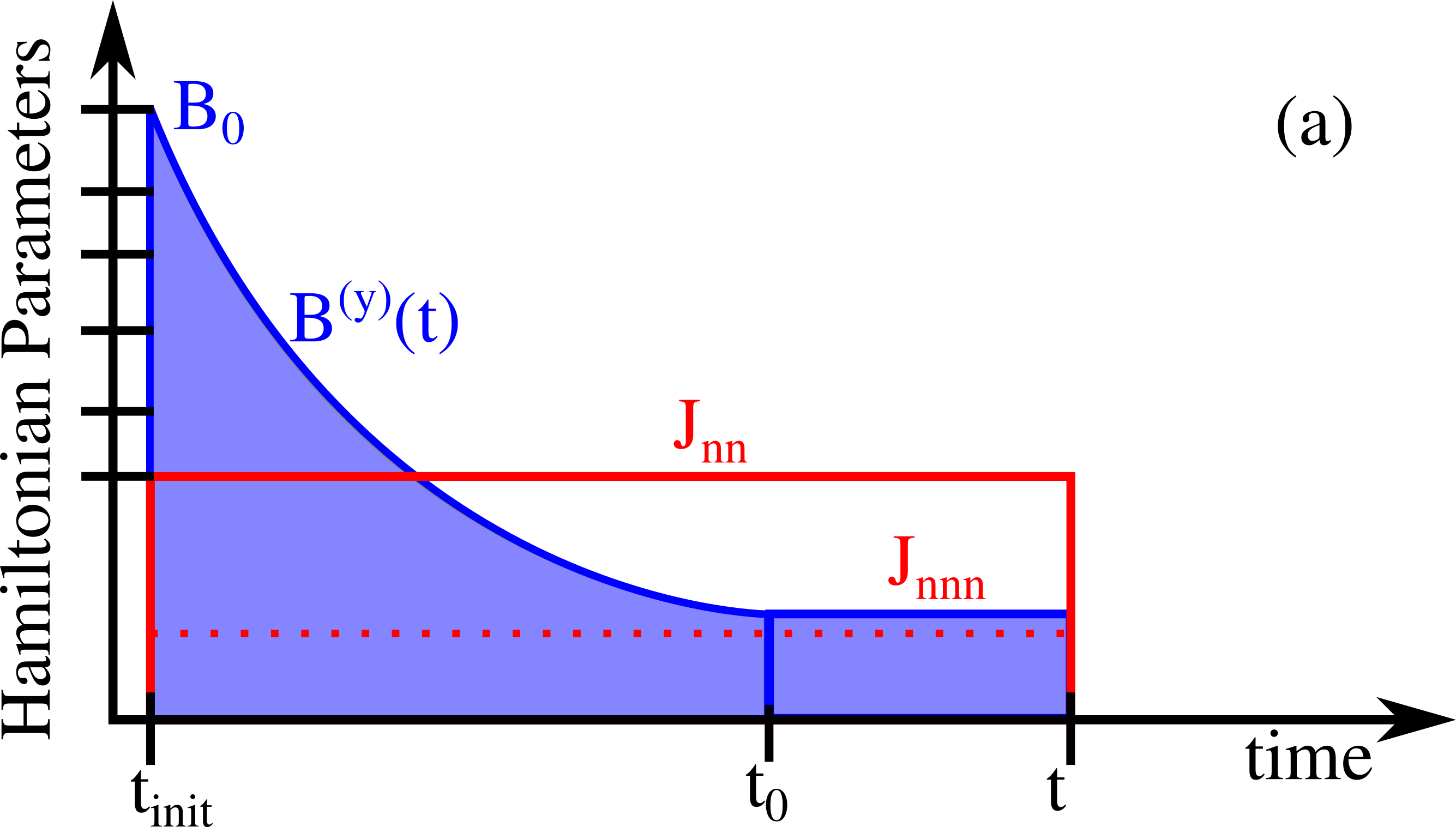}\\
		\includegraphics[scale=0.27]{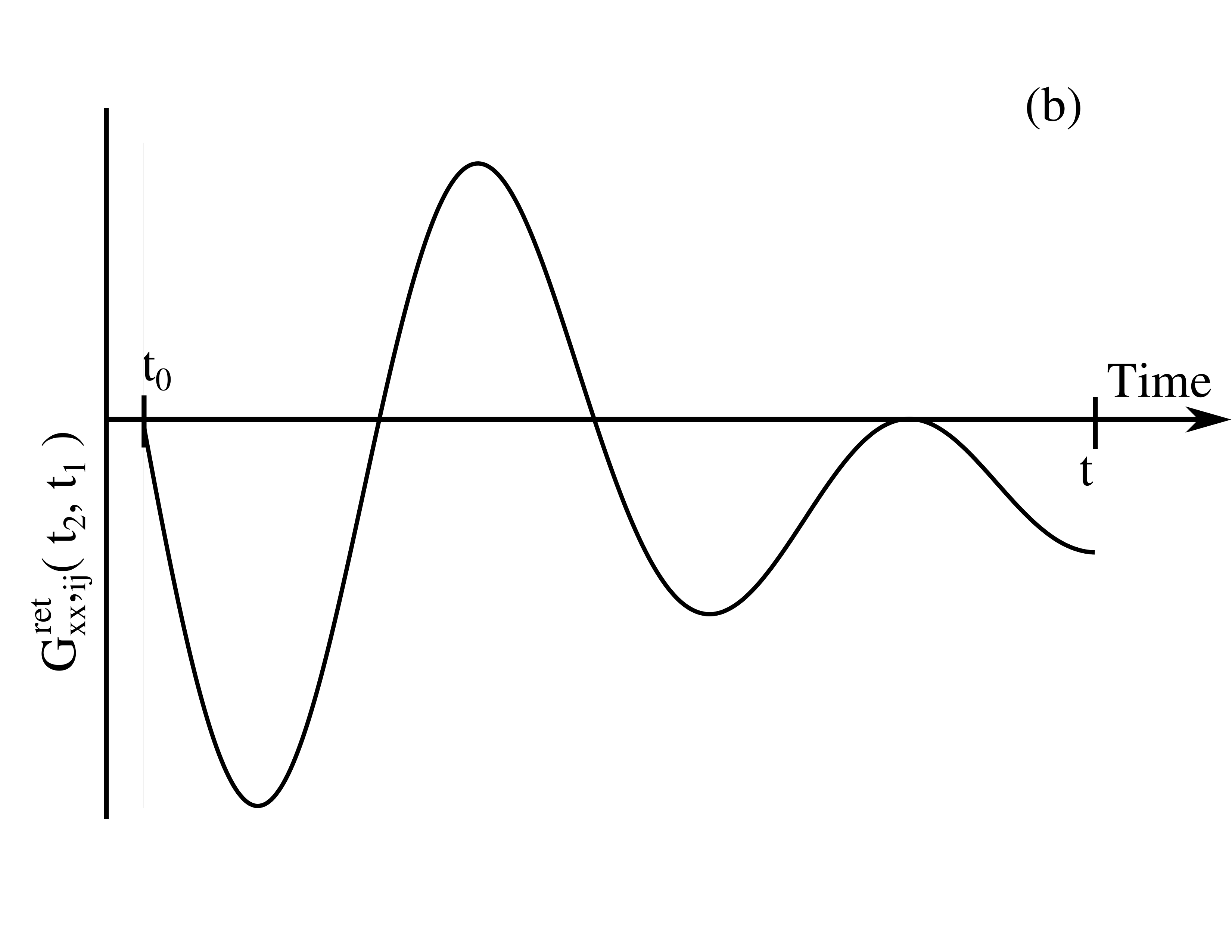}
	\end{tabular}
	\caption{(Color online.) Schematic of the complete protocol we use in implementing the Ramsey spectroscopy. (a) We initialize the state in the ground state of the Hamiltonian when $B^{(y)}\gg |J_{ij}|$. Then we decrease the transverse magnetic field via an exponential ramp in time. (b) For the time interval $[t_0, t]$, we apply the Ramsey spectroscopy protocol  and perform signal processing on the resulting measurements as a function of the final time $t$.}
	\label{fig:Protocol}

\end{figure}

When we examine the spin-spin Green's function, we can consider a number of different scenarios. We start the system at $t_{init}$ in the ground state of the Hamiltonian when $B^{(y)}(t_{init})\gg |J_{ij}|$, evolve it by decreasing the field to time $t_0$, then apply the Ramsey protocol. During the Ramsey spectroscopy, we can keep the transverse field constant (as we will do here) or we can continue to vary it in time until $t$, when the final measurement is made.  While we focus on the case when the Hamiltonian is a constant during the Ramsey protocol in this work, the more general case allows one to investigate strong nonequilibrium effects associated with the spin-spin Green's function. Unfortunately, there is no simple way to interpret the results of those such experiments, which is why we focus on the simpler case here, which can be directly interpreted. The protocol is illustrated in Fig.~\ref{fig:Protocol}.

The formula for the $J_{ij}$ have been derived in Ref.~\cite{Zhu} and the resulting equation for the static $J_{ij}$ is
\begin{equation}
J_{ij} = \Omega\nu_R \sum_{m=1}^N \frac{b_{im} b_{jm}}{ \mu^2 - \omega^2_m}.
\label{eq:Jij}
\end{equation}
The $J_{ij}$ depend on the normal mode eigenvector, $b_{im}$, of the $m^{\text th}$ phonon mode at the $i^{\text th}$ ion site and the corresponding phonon frequency, $\omega_{m}$ --- the calculation of $b_{im}$ and $\omega_{m}$ can be found in Ref.~\cite{James}. The remaining variables in Eq.~(\ref{eq:Jij}) are experimental parameters and we use the same parameters used in Ref.~\cite{science} (we work with conventional frequency units). The symbol $\nu_R = h/(M\lambda) = 18.5$ kHz is the recoil energy of a $^{171}$Yb$^+$, where $M$ is the ionic mass, $\lambda = 355$ nm  is the wavelength of the laser applied to the linear chain of ions, and $\Omega = 600$ kHz is the Rabi frequency. The parameter $\mu$ is the detuning and is defined by $\mu = \omega_{COM} + 3\eta\Omega= 1.0233\omega_{COM}$, where $\omega_{COM}$ is the transverse center of mass phonon mode and the Lambe-Dicke parameter $\eta = \sqrt{ \nu_R/ \omega_{com}}=0.0621$. The $J_{ij}$ can be adjusted to yield different power law behavior, as described in Eq.~(\ref{eq:approx}), by changing the detuning, $\mu$, or the asymmetry between the axial and transverse center of mass mode. We use the latter strategy in Sec. III. The axial center of mass mode is adjusted from $620$ kHz to $950$ kHz yielding a power law fit ranging from $0.7 < \alpha < 1.2$. The energy unit, $J_0$, that we use to scale the $J_{ij}$ satisfies $J_0 \approx 1$ kHz for $N = 10$ ions. In Sec. III, we focus on the ferromagnetic interaction  with positive spin-exchange coefficients ($J_{ij} > 0 $). 

The specifics of the Ramsey protocol we use is as follows:
 (1) Initialize the system of spins along the $y$-direction at $t_{init}$ and start with $B^{(y)}(t_{init})=B_0=10J_0$, (2) reduce the transverse magnetic field via an exponential ramp with $B^{(y)}(t)=B_0 \exp(-t/\tau)$ between $t_{init}$ to $t_0$, (3) apply the Ramsey protocol in the time interval $[t_0, t]$ with a constant transverse magnetic field, $B^{(y)}(t_0)$ and (4) perform a signal processing analysis on the resulting measurements. The Ramsey protocol is the same as described above with the single spin rotation at $t_0$ and the pure state $|\psi_0\rangle$ is the state that was time evolved from $t_{init}$ to $t_0$. 
Note that during step 2, the transverse magnetic field changes in time, requiring a time-ordered-product 
for the evolution operator until time $t_0$.

\subsection{Time Evolution }
We must evaluate the time evolution with respect to the time-dependent Hamiltonian that satisfies
\begin{equation}
	i\frac{\partial}{\partial t} \hat{U}(t, t_{init}) = \hat{\mathcal{H}}(t) \hat{U}(t,t_{init}),
	\label{eq:TimeEvolve}
\end{equation}
where $\hat{U}(t_{init}, t_{init}) = \hat I$. The resulting time evolution operator is a time-ordered product, $\hat{U}(t, t_{init}) = \mathcal{T}_t \exp\left[-i \int_{t_{init}}^t \mathrm{d}t'\hat{\mathcal{H}}(t')\right]$. After $t_0$, the magnetic field is held constant and the transverse-field Ising model is time independent, which simplifies the subsequent time evolution operator to $\hat U(t,t_0)| n\rangle = \exp[ -i E_n (t - t_0 )]| n\rangle$ for the eigenstates defined above at $t_0$.

We follow the same procedure as we did in Ref.~\cite{us_spectroscopy} and use the commutator-free exponential time (CFET) approach to approximate the nontrivial time evolution operator. The details of the CFET approach can be found in Refs.~\cite{cfet1, cfet2}. The central idea of the CFET approach is to use a number of Trotter approximations to construct a single evolution operator that evolves a $\delta t$ forward in time. The Trotter factors are chosen in a manner that when combined via the Baker-Campbell-Hausdorff formula~\cite{Baker, Campbell, Hausdorff} they produce a high-order truncated Magnus expansion~\cite{magnus} of the evolution operator. Depending on the number of Trotter factors used, the CFET operator can increase the order of the truncated Magnus expansion. We use the optimized fourth-ordered CFET approach, that has an error of $\delta t^5$.

\subsection{ The spectral function and spectral moments }

The spectral function determines the local density of states of the quantum system. Spectral moment sum rules are useful to understand the short time behavior and can ultimately be applied to yield Lieb-Robinson bounds. While the spectral moment sum rules can be derived for the completely general nonequilibrium Green's function, we do so only for the case of a Hamiltonian that is constant for times $t>t_0$ here.

The spectral function is then defined via
\begin{equation}
	\begin{aligned}
	&A_{xx,ij}^{ret}( \omega) = \\
	 & - \frac{1}{\pi} \text{Im}\left[ \int \limits_{0}^{\infty} \mathrm{d}t_{rel}G_{xx,ij}^{ret}\left(t_{0} + t_{rel}, t_{0} \right) \mathrm{e}^{i\omega t_{rel}}\right]
	\end{aligned}
\end{equation}
and the $n^\text{th}$ spectral moment is then defined as follows:
\begin{equation}
	\mu_{xx, ij}^{ret,n} = \int \limits_{-\infty}^{\infty} \mathrm{d}\omega \omega^n A_{xx,ij}^{ret}( \omega).
\end{equation}
Using integration by parts $n$ times, one can directly relate the $n^\text{th}$ spectral moment to the $n^\text{th}$ derivative of the retarded Green's function as 
\begin{equation}
	\begin{aligned}
	&\mu_{xx,ij}^{ret, n} = \\
	&-\frac{2}{\pi}\text{Im}\left[ i^n \frac{\partial^n }{\partial t_{rel}^n} G_{xx,ij}^{ret}\left(t_0 + t_{rel}, t_{0} \right)\right]_{t_{rel} = 0^+}.
	\end{aligned}
\end{equation}
We calculate the first nonzero spectral moments for arbitrary lattice sites $i$ and $j$. The calculations are tedious, but straightforward and finally yield
\begin{widetext}
\begin{subequations}
\begin{align}
&\mu_{xx,ij}^{ret, 0} = 0, \\
&\mu_{xx,ij}^{ret, 1} = \frac{4}{\pi} B^{(y)}(t_{0}) \langle \psi_0 | \sigma^{(y)}_i | \psi_0 \rangle \delta_{ij}, \\
&\mu_{xx,ij}^{ret, 2} = 0, \\
&\mu_{xx,ij}^{ret, 3} = -\frac{ 8 }{\pi} \left( B^{(y)}(t_{0}) \right)^2 J_{ij}\langle\psi_0|\sigma_i^{(y)} \sigma_j^{(y)}|\psi_0\rangle + \frac{\delta_{ij} }{2\pi} \langle \psi_0|\left[ B^{(y)}(t_{0}) \left( 16 \left( B^{(y)}(t_{0}) \right)^2 + \sum_{k k'} J_{ik}J_{ik'}\sigma_k^{(x)} \sigma_{k'}^{(x)} \right) \right]\sigma_i^{(y)}|\psi_0\rangle \nonumber\\
& + \frac{4 \delta_{ij} }{\pi } \sum_k J_{ik} \langle \psi_0|\left [\left( B^{(y)}(t_{0}) \right)^2 \left( \sigma_k^{(x)} \sigma_i^{(x)} + \sigma_k^{(z)} \sigma_i^{(z)}\right)  \right]|\psi_0\rangle.
\end{align}
\label{eq:moments}
\end{subequations}
\end{widetext} 
The zeroth and first two moments vanish except when $i=j$, where the first moment is nonzero. This implies the Green's function for $i\ne j$ is very flat in $t$ initially; the case with $i=j$ has a nonzero slope that is proportional to the transverse magnetic field and the polarization of the spin along the field direction. For $i\neq j$ the first nonzero spectral moment is $\mu_{xx,ij}^{ret, 3}(t_0 )$ and here the coefficient is proportional to the direct spin-spin interaction between $i$ and $j$ and rotated from along $x$ to along $y$---the coefficient is scaled also by the square of the transverse magnetic field. The case with $i=j$ is even more complicated.

\subsection{Signal processing}

The generalized Lehmann formula in  Eq.~(\ref{eq:lehmann}) shows that the time dependence of the Green's function relates to
the different excitation energies of the many-body system. Since these excitation energies are discrete,
the time dependence is determined by a finite set of exponentials with different weights.  This is precisely
the case where compressive sensing can be employed to extract the frequencies and the weights most efficiently from the data in the time domain; this becomes particularly important since the extent of the time domain is limited by the decoherence time in an experiment.
Even though compressive sensing is optimized for this procedure, it remains numerically challenging because there are a fairly large number of nonzero frequencies and some of them lie close to one another, so good data is necessary to extract all of them; instead, we use a small number of points to simulate the case of an actual experiment and hence we won't be able to pick out all of the frequencies.
A complete review of compressive sensing for clean and noisy signals can be found at Rice University~\cite{rice}. 

To Fourier transform a data set of $M$ time steps to a frequency domain with $N_{step}$ frequency steps, the following matrix-vector multiplication equation is solved
\begin{equation}
	A\mathcal{F}^{-1}G_{xx,ij}^{ret}( \omega_n) = AG_{xx,ij}^{ret}( t_m, t_0 ).
	\label{eq:solve_this}
\end{equation}
Here, the inverse Fourier transform matrix, $\mathcal{F}^{-1}$, is performed by a partial discrete Fourier transform, where the matrix size is $M\times N_{step}$, $A$ is the measurement matrix that is of size $M\times M$, and the index $n$ runs over the $N_{step}$ frequency steps, while the index $m$ runs over the $M$ time steps. The inverse Fourier transform matrix satisfies
\begin{equation}
\mathcal{F}^{-1}_{mn}=\frac{1}{N_{step}}e^{-i\omega_n t_m},
\end{equation}
The construction of the measurement matrix, $A$, is one of the key elements of compressive sensing. It is a random orthogonal matrix, whose elements, $A_{ij}$, are randomly chosen from a normal distribution and then $A$ is orthogonalized, where the columns of the matrix are the vectors. When Eq.~(\ref{eq:solve_this}) is solved, there are an infinite number of possible solutions, due to the fact that Eq.~(\ref{eq:solve_this}) is an underdetermined system of equations. The other tenet of compressive sensing is that the solution that minimizes the absolute value of the signal in the frequency domain  
\begin{equation}
	\text{min } ||G_{xx,ij}^{\rm ret} \left(  \omega \right) ||= \text{min} \sum_n
|G_{xx,ij}^{\rm ret} \left(  \omega_n \right) |
	\label{eq:minimize} 
\end{equation}
is the optimal solution to pick from the different choices. In addition, one can randomly choose the time 
coordinates (instead of having them on a uniform grid), but we do not use this additional randomness in this work. We employ the matlab toolkit CVX~\cite{CVX} to solve Eq.~(\ref{eq:solve_this}) subject to the constraint in Eq.~(\ref{eq:minimize}).   

\begin{figure}
	\centering
	\begin{tabular}{ c }
		\includegraphics[scale=0.065]{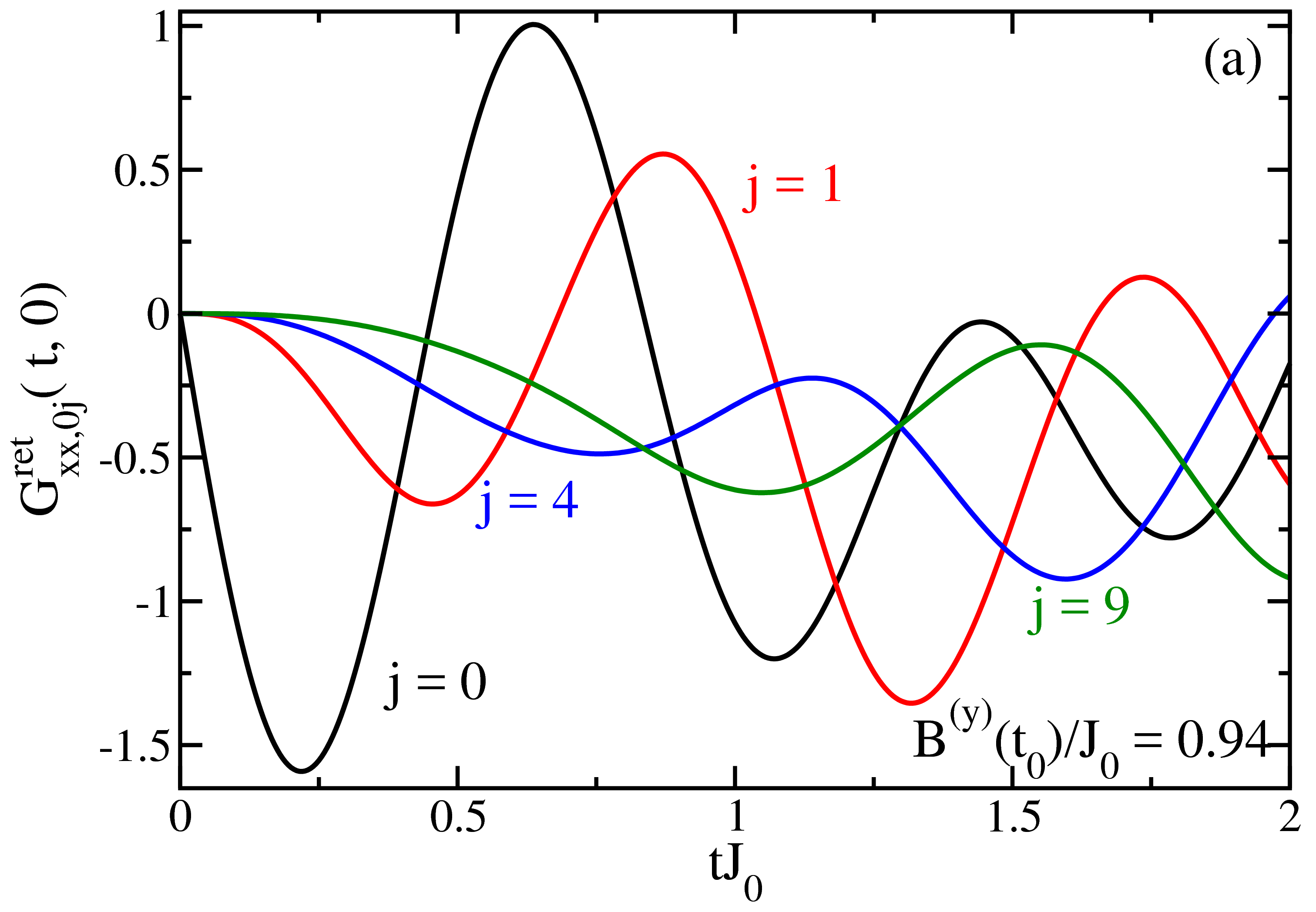} \\
		\includegraphics[scale=0.065]{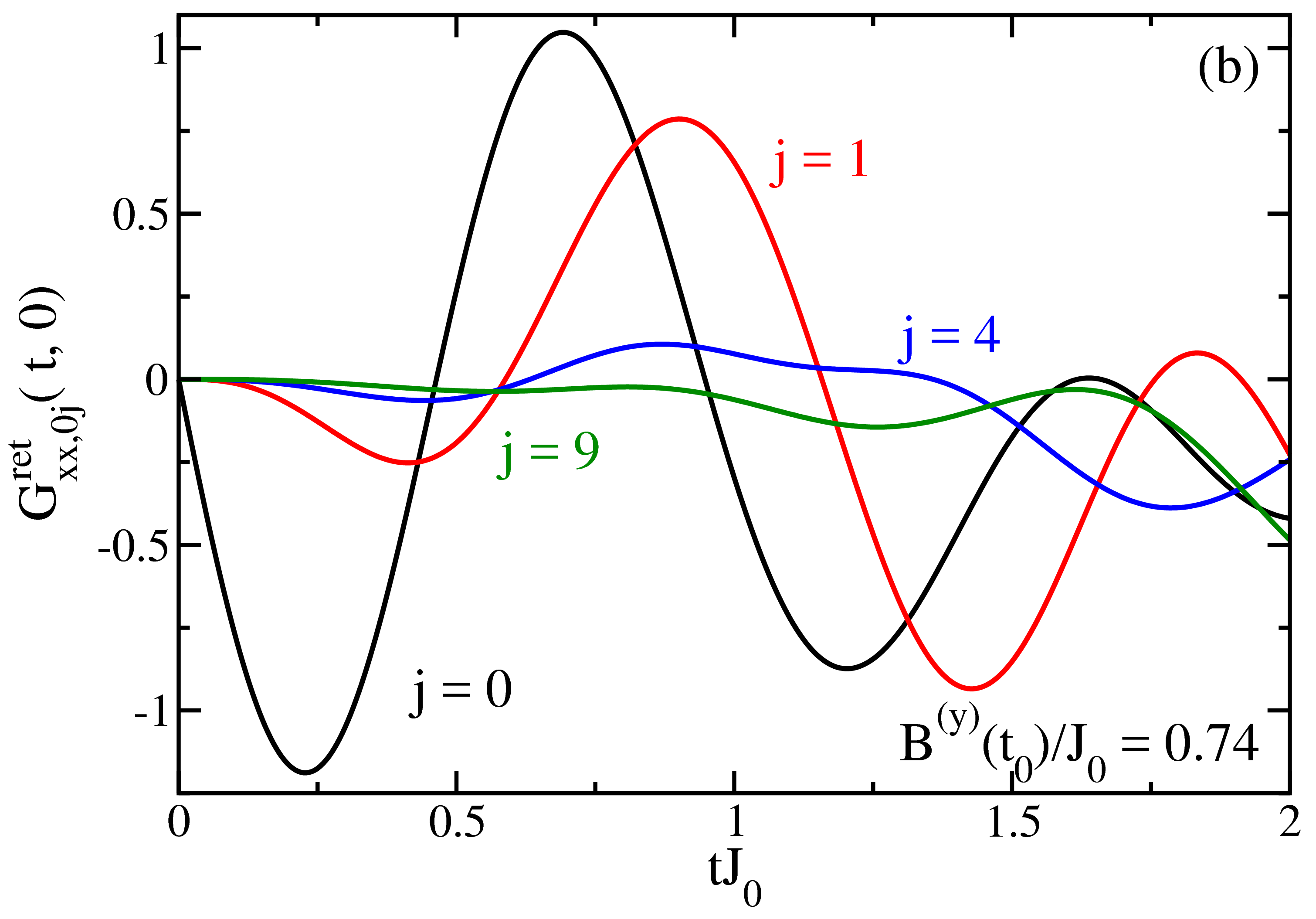}\\
		\includegraphics[scale=0.065]{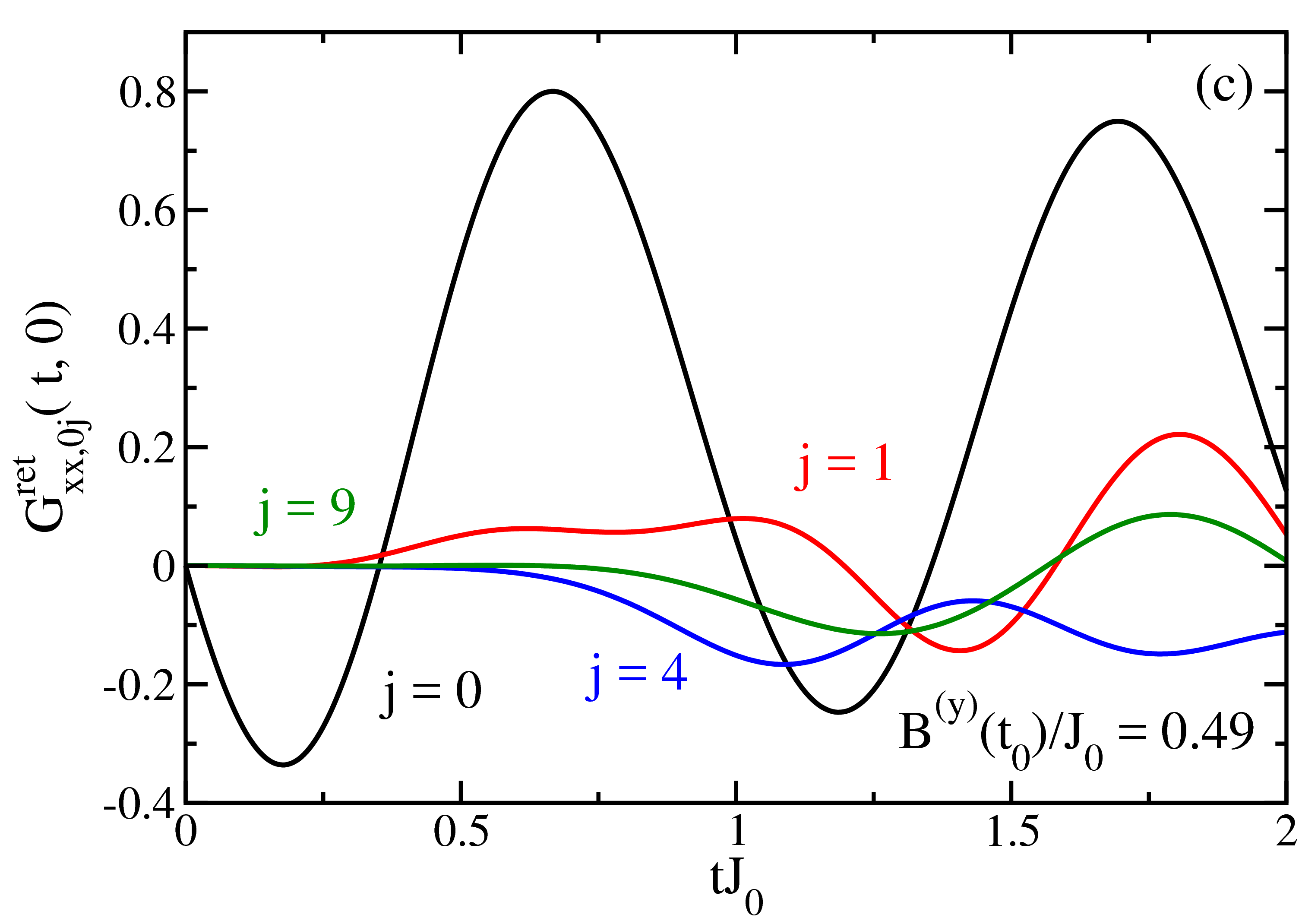}\\
		\includegraphics[scale=0.065]{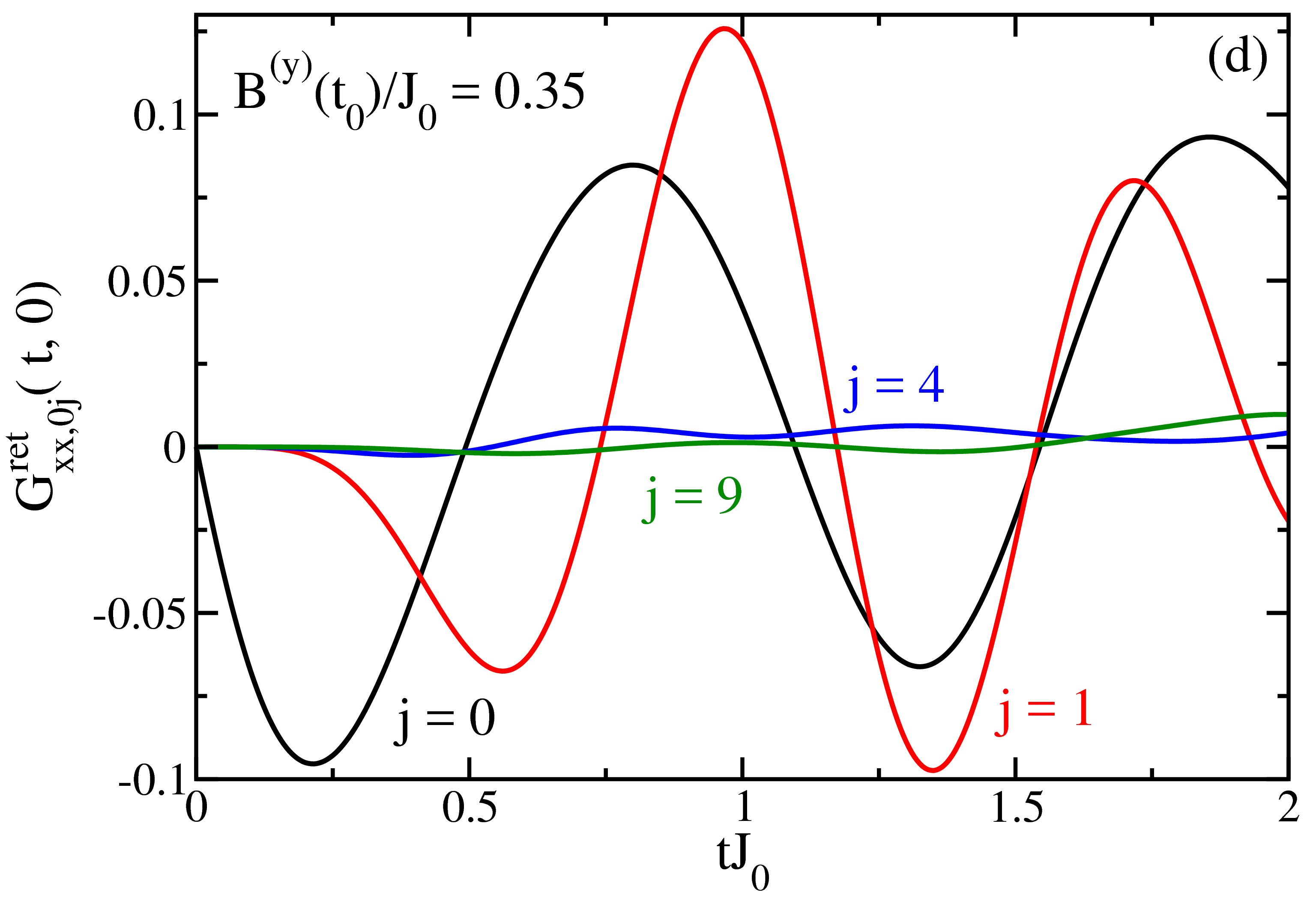}
	\end{tabular}
	\caption{(Color online.) Examples of the Ramsey spectroscopy as a function of time for $i=0$ and $j = 0$ (black), $1$ (red), $4$ (blue), and $9$ (green) with $4$ different $B^{(y)}(t_0)/J_0$ values [$B^{(y)}(t_0)/J_0$ is equal to the following: (a) $0.94$, (b) $0.74$, (c) $0.49$, and (d) $ 0.35$]. }
	\label{fig:TimeSlice}
\end{figure}

\section{Results} 

We show numerical examples of the application of the Ramsey spectroscopy protocol to a linear chain of $N=10$ spins with a ferromagnetic interaction, $J_{ij} > 0$. We perform the single spin rotation on the spin at the left end of the linear chain, $i = 0$, and the characteristic transverse magnetic field ramping time is $\tau = 0.85/ J_0$.   

We show examples of the Ramsey spectroscopy protocol as a function of time for four pairs of spins, the first spin is always site $i=0$ and the second spin is $j = 0$, $1$, $4$, and $9$ (for the $N=10$ spin chain). We study four different final transverse magnetic fields, $B^{(y)}(t_0)/J_0 = 0.94$, $0.74$, $0.49$, and $ 0.35$. From this point on, we will assume $t_0=0$ to make the discussion simpler. As expected from Eq.~(\ref{eq:moments}), the Green's function for all spins at $t_2 = t_0=0$ starts at $0$. A little afterwards, the $G_{xx,00}^{\rm ret}(t,0)$ decreases linearly as a function of time, and the slope becomes more shallow as the transverse magnetic field decreases, also as expected from the sum rules. Additionally, at the three other sites, $j = 1$, $4$, and $9$ the Green's function begins with a very flat $t$ dependence, because the first two derivatives vanish.  and then decreases at different rates dependent on the distance of the spin from the left edge $i=0$ of the lattice. Note, that as the transverse magnetic field decreases the amplitude of the measurements also decreases. The decrease in the amplitude is due the eigenstates of the transverse-field Ising model becoming the eigenstates of $\sigma^{(x)}$ as  $B^{(y)}(t_0) \xrightarrow {} 0$. 

The oscillations for $i = j = 0$ can be interpreted in an alternative manner via a Loschmidt echo~\cite{loschmidt,Knap}. The Loschmidt echo describes a forward propagation in time with one Hamiltonian $\mathcal{H} = \mathcal{H}_0 +V$, and then a backward propagation in time with another Hamiltonian $\mathcal{H}_0$, 
\begin{equation}
	\mathcal{L}(t-t_0) = \langle \psi | \text{e}^{ i \mathcal{H}_0 (t-t_0)}\text{e}^{ -i \mathcal{H} (t-t_0)} | \psi \rangle.
\end{equation}
The pure state Green's function in Eq.~(\ref{eq:Greens}) can then be defined in terms of a Loschmidt echo by realizing that $ \sigma_i^x \exp^{ -i \mathcal{H}_0 (t-t_0)}\sigma_i^x = \exp^{ -i \mathcal{H} (t-t_0)}$, so that the local pure state retarded Green's function is rewritten as
\begin{equation}
	G_{x x , ii}^{\rm ret}(t,t_0) = - i \theta( t- t_0) \left[ \mathcal{L}(t-t_0) - \mathcal{L}(-t+t_0) \right].
\end{equation}
We examined the Loschmidt echo time trace over a long time interval given by a length of 100~ms. The
Green's function has significant oscillations here, sometimes including low frequency oscillations with large amplitudes (not shown here). By comparing the features at different times by eye, we notice that the amplitude of the oscillations appears to remain large and not decay exponentially. 
This identifies that this spin system appears to be in the localized regime because the oscillations do not seem to decay as a function of time (if they did decay, then it would be in the diffusive regime). 

\subsection{ Lieb-Robinson bounds}
\begin{figure*}[!ht]
	\centering
	\begin{tabular}{ c c }
		\includegraphics[scale=0.065]{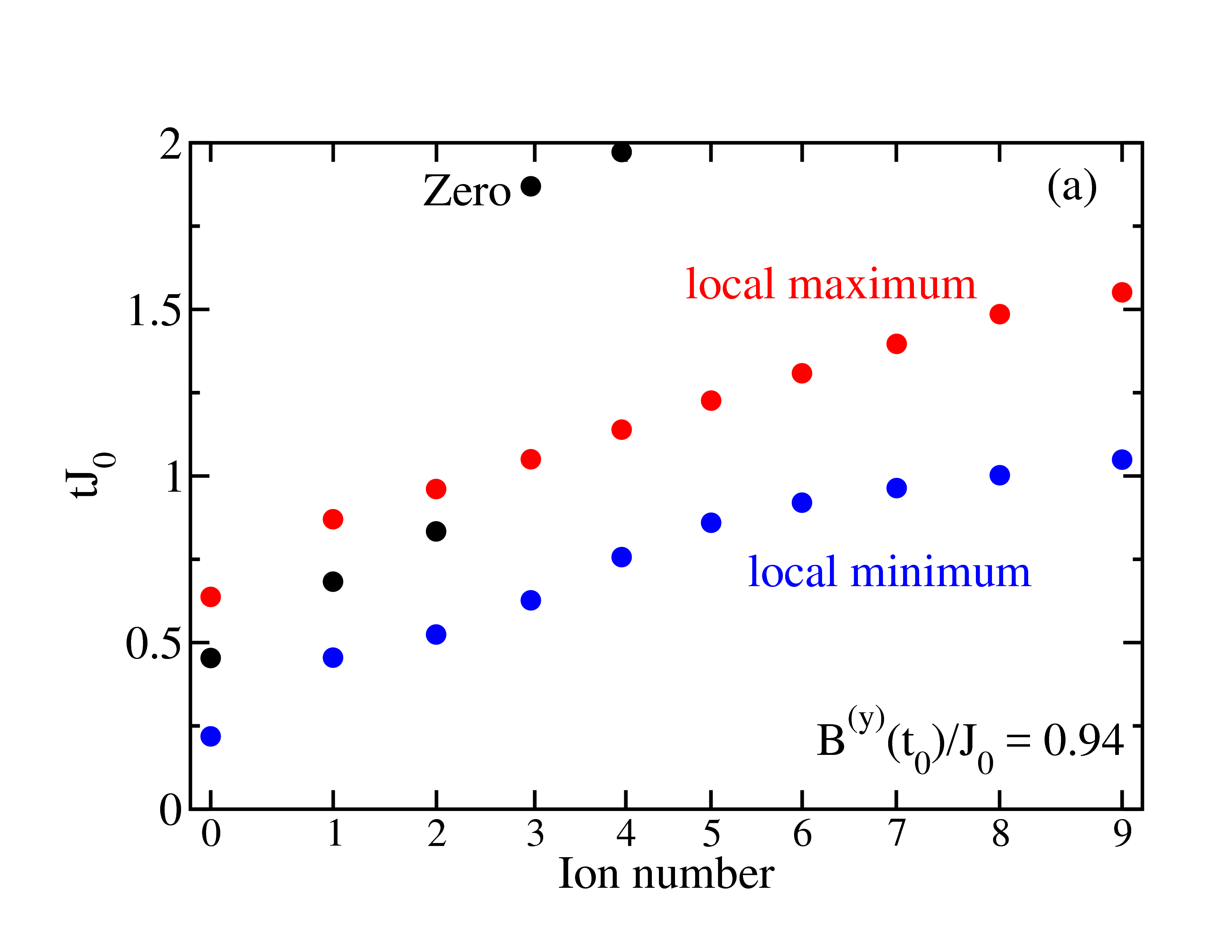} & \includegraphics[scale=0.065]{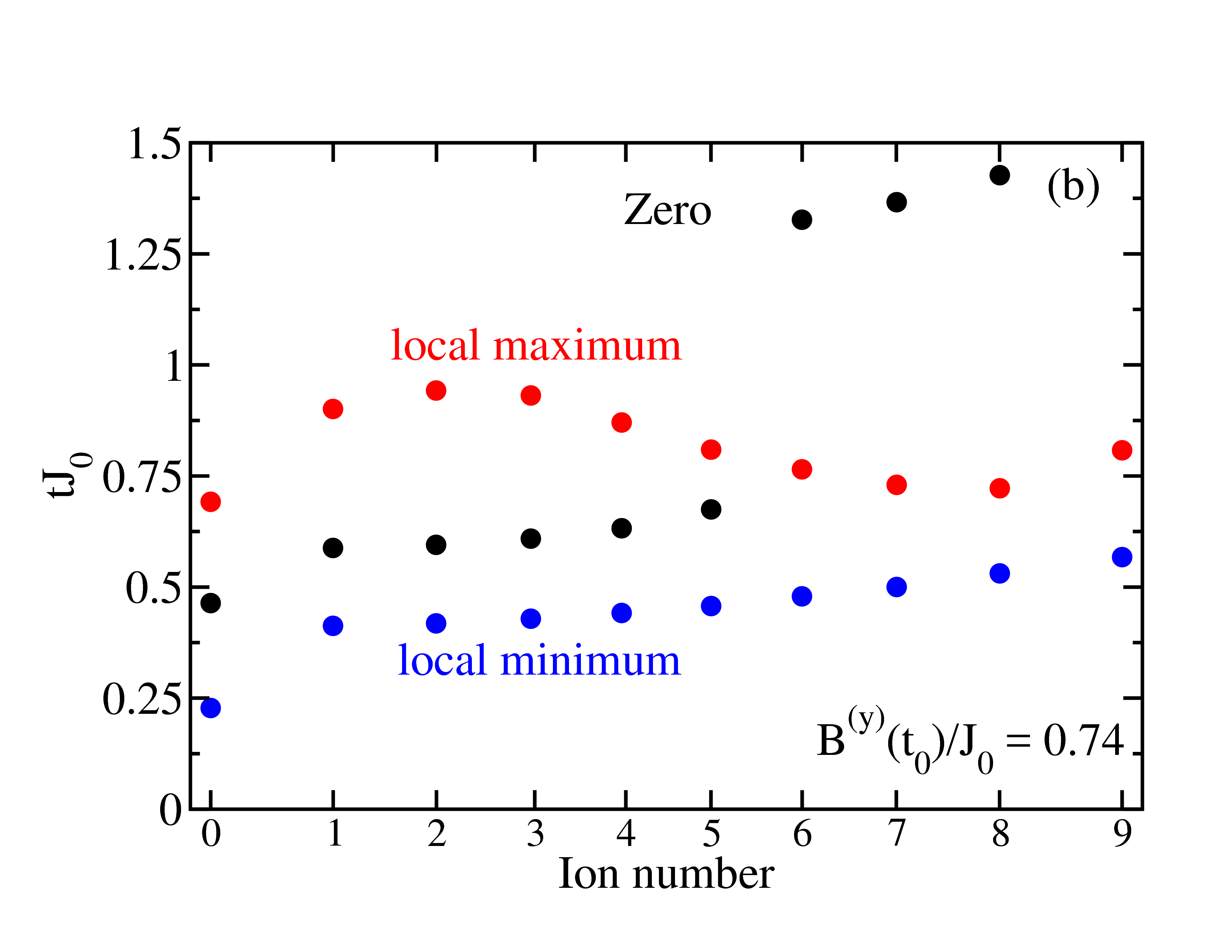}\\
		\includegraphics[scale=0.065]{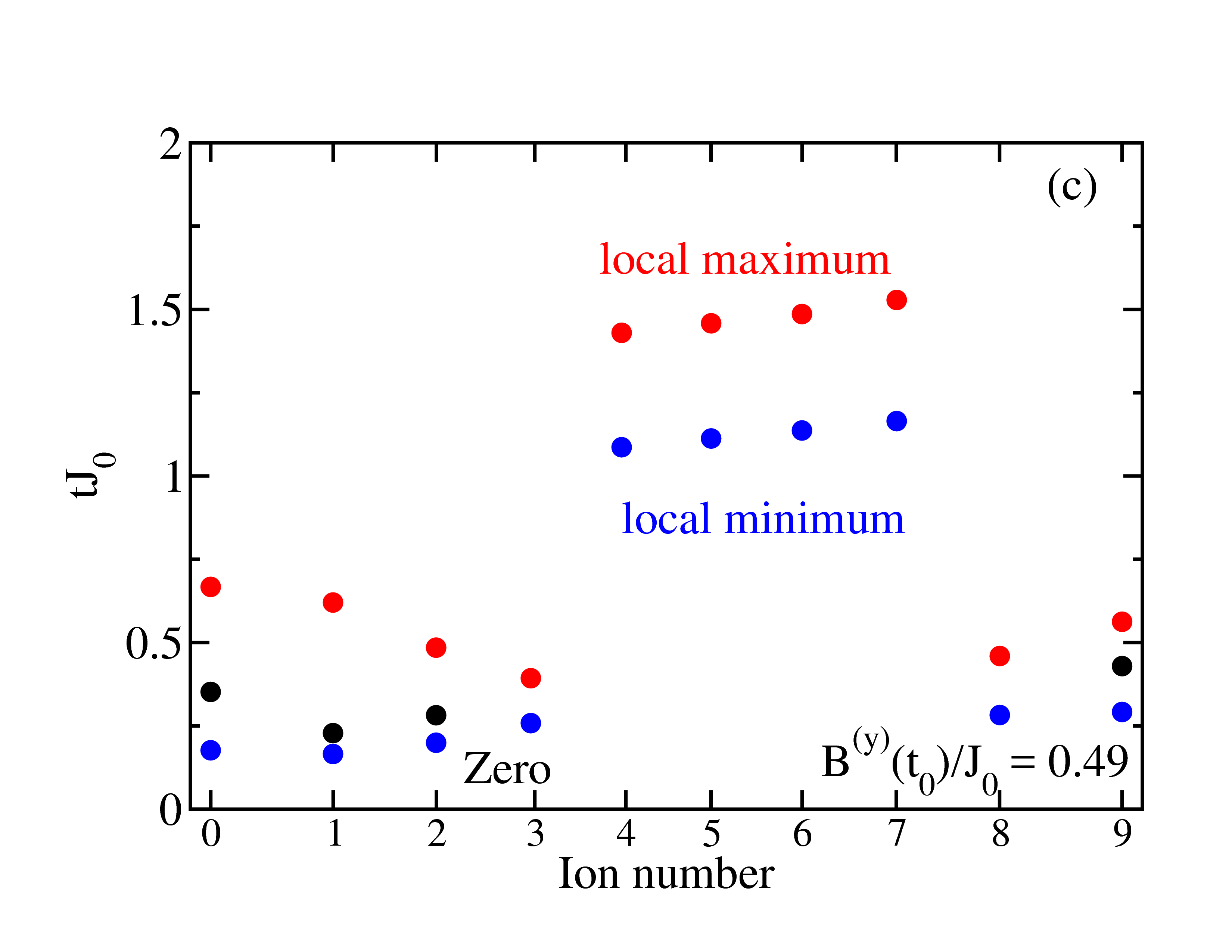} & \includegraphics[scale=0.065]{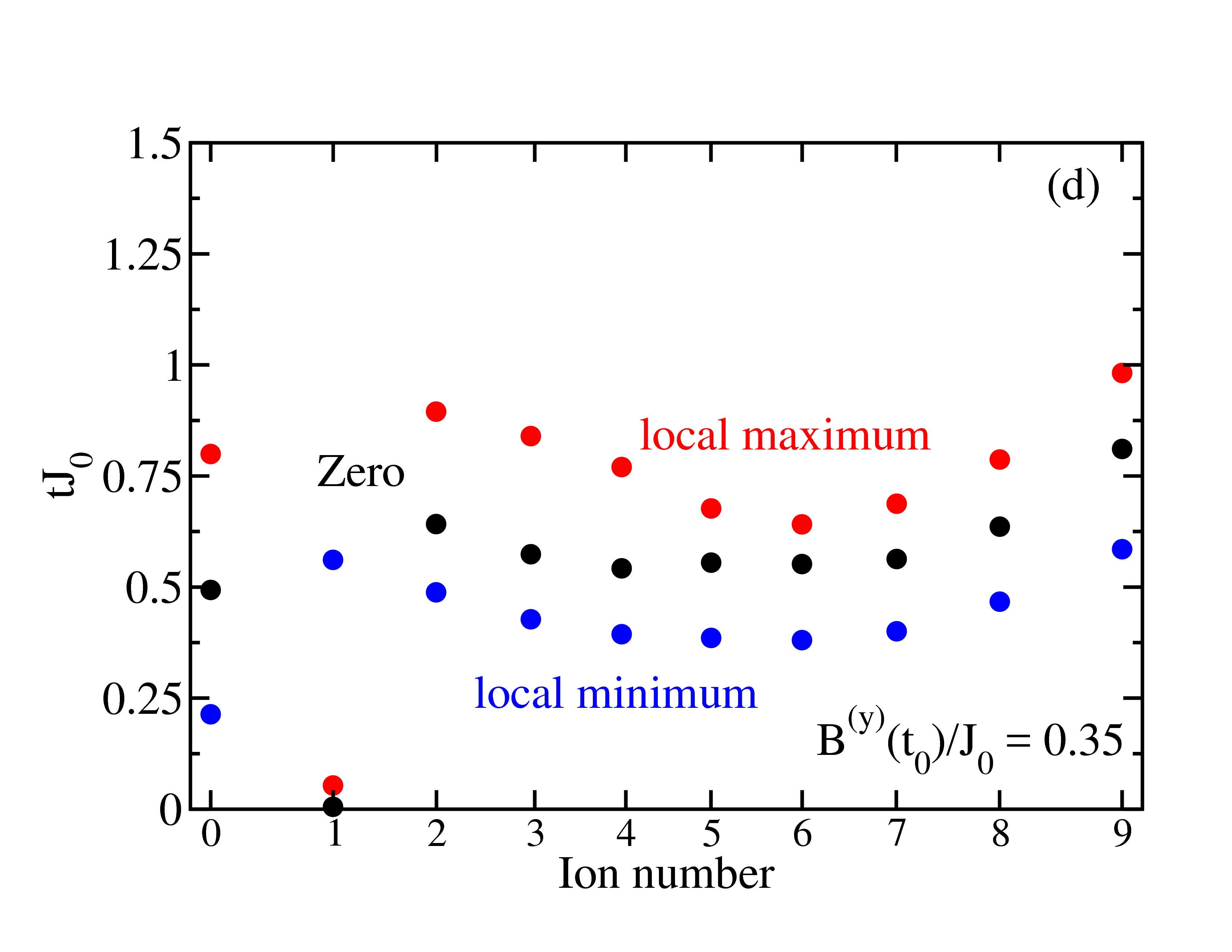}
	\end{tabular}
	\caption{(Color online.) First local minimum (blue circles), local maximum (red circles) and zero (black circles) of the pure state Green's function for $i=0$ and all other $j$ lattice sites. Here the $x$-axis is labeled at the equilibrium position from the left edge $i = 0$, $|R_0 - R_j|$. }
	\label{fig:Alternatives}

\end{figure*}

Next, we want to identify how the pure state Green's function can be employed to examine Lieb-Robinson-like behavior. Here, we have a system that has a perturbation initiated at the left end of the chain, and we can ask how long does it take for the perturbation to be seen elsewhere in the chain. Since the system has long-range interactions, we expect the information to flow with a power-law behavior rather than a light-cone, as determined recently~\cite{hastings,gorshkov}. One idea to track this information flow is to track some feature of the Green's function which measures the time-delay for the response. In looking at the results in Fig.~\ref{fig:TimeSlice}, we see that the first minimum, first maximum, and first zero crossing all seem to correlate with the distance from the left end of the chain. So we plot the times at which those features occur for the different lattice sites in Fig.~\ref{fig:Alternatives}, with each panel corresponding to a different final transverse field:
$B^{(y)}(t_0)/J_0 = 0.94$, $0.74$, $0.49$, and $ 0.35$.   

In Fig.~\ref{fig:Alternatives}~(a), the first local minimum and maximum seemingly have a power law behavior as a function of the relative distance from $i = 0$, while the first zero is only observed for spins $j < 5$. As the transverse magnetic field, in Fig.~\ref{fig:Alternatives}~(b-d), is decreased, the power law behavior of the first local minimum and maximum becomes unrecognizable. The first zero crossing in Fig.~\ref{fig:Alternatives}~(c) seemingly jumps from $j = 2$ to $j = 9$. The behavior of these features at the low transverse magnetic field are not useful in investigating Lieb-Robinson bounds. They show no regular behavior and hence must be governed by other physical phenomena.

\begin{figure*}[!ht]
	\centering
	\begin{tabular}{ c c }
		\includegraphics[scale=0.065]{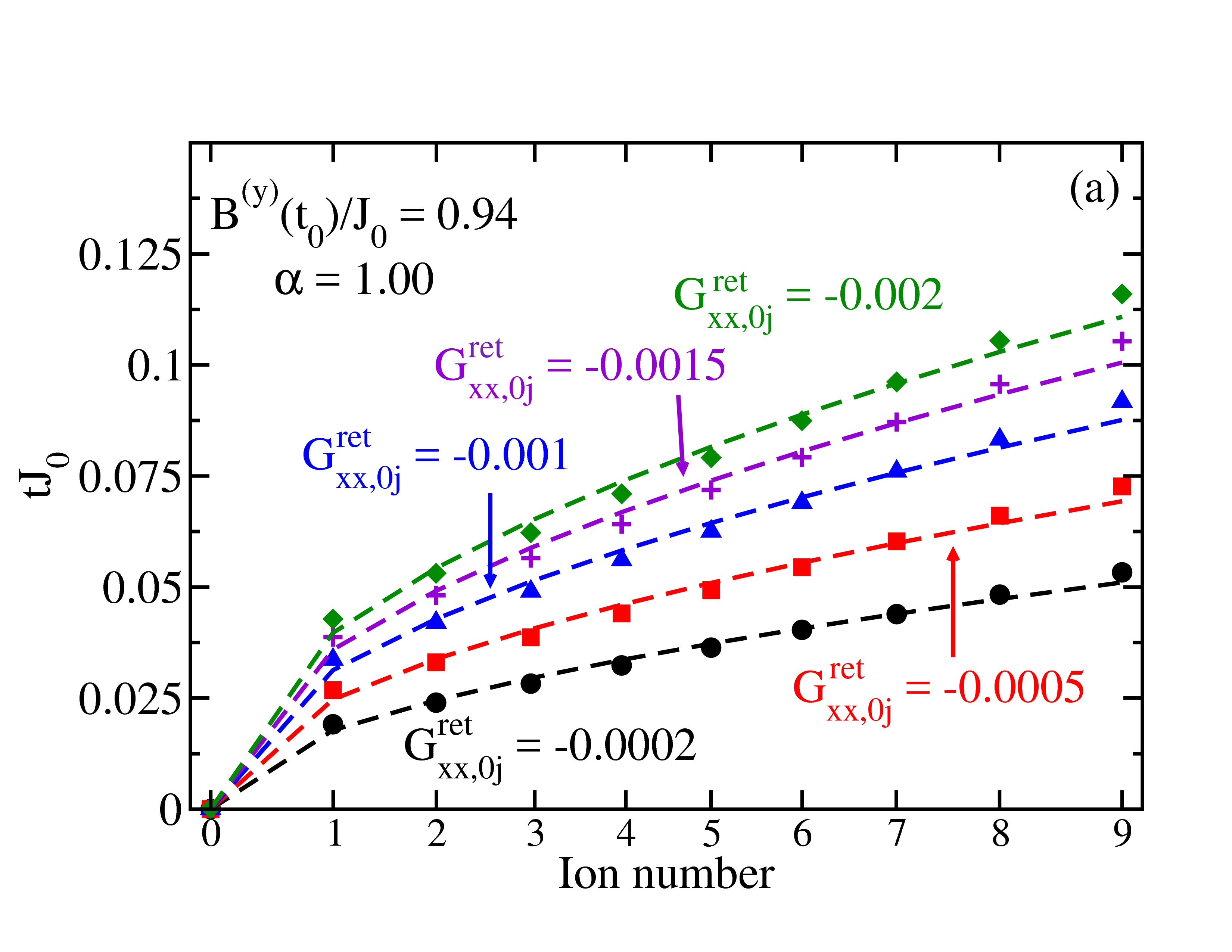} & \includegraphics[scale=0.065]{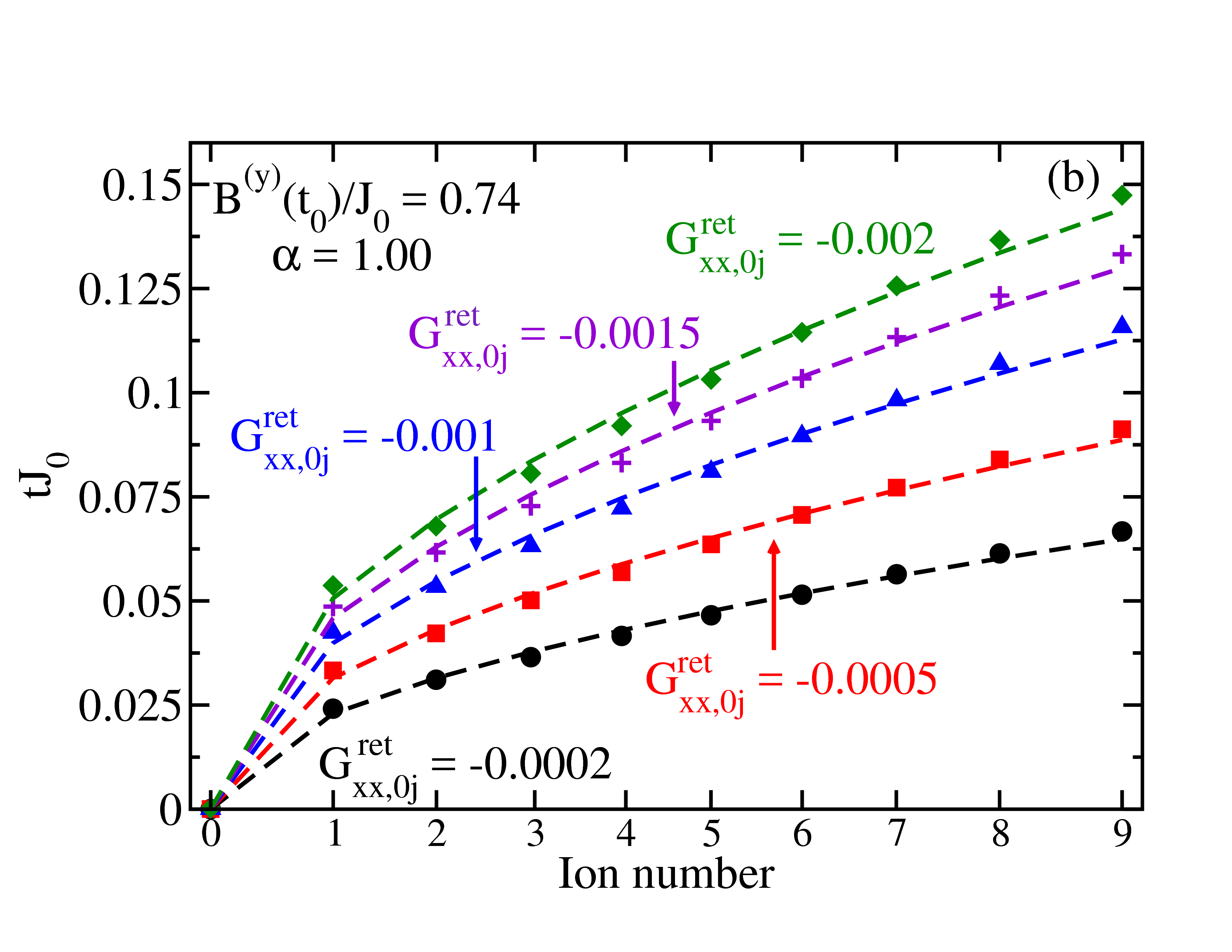}\\
		\includegraphics[scale=0.065]{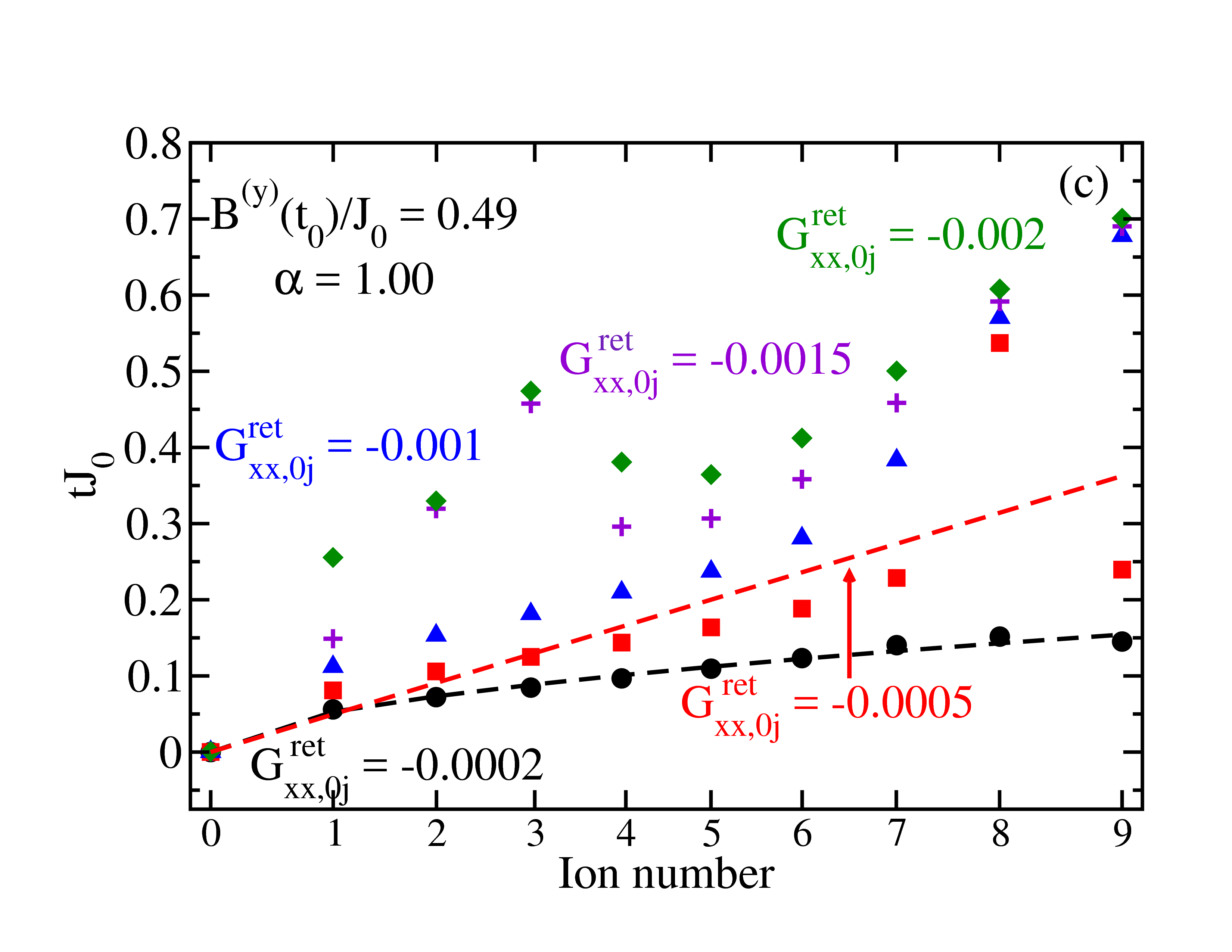} & \includegraphics[scale=0.065]{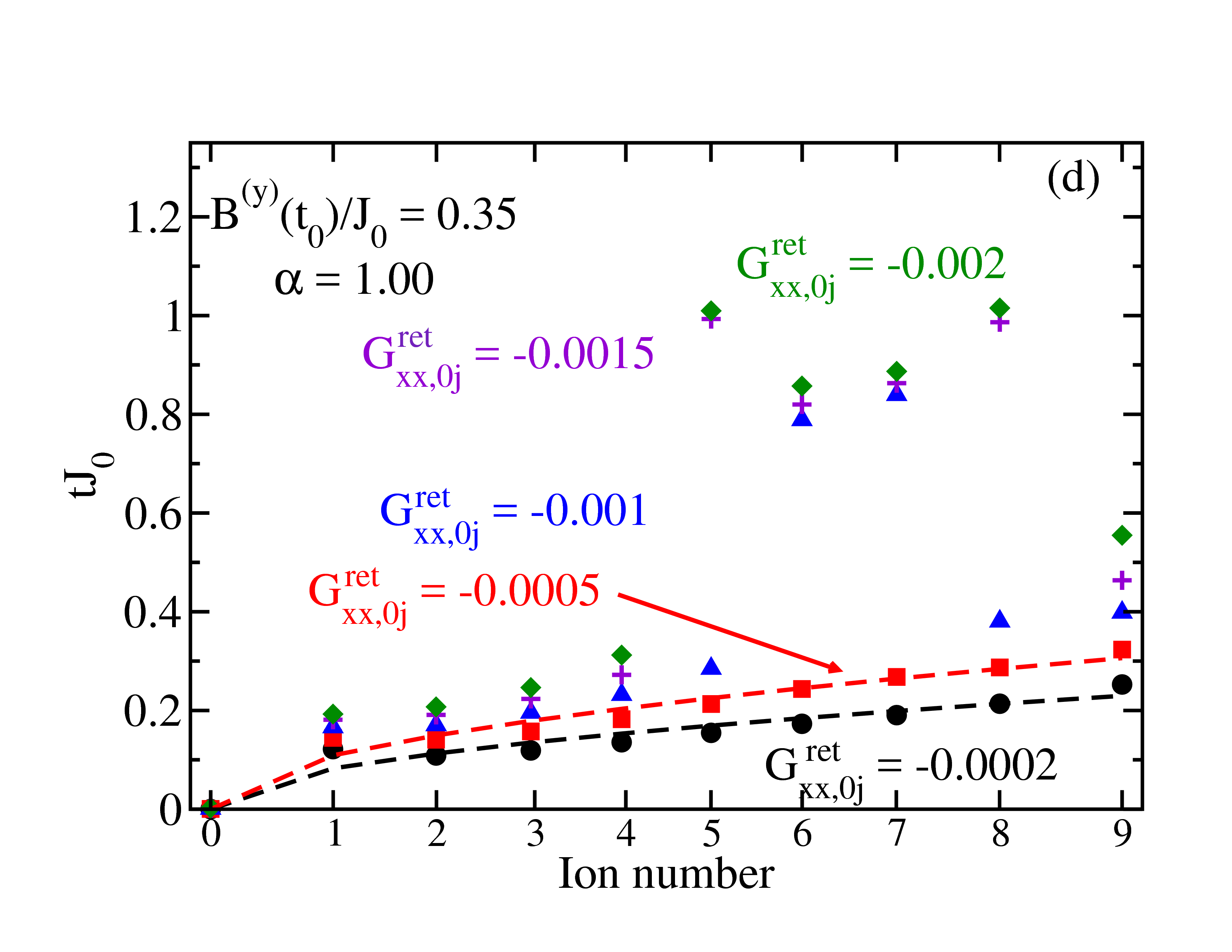}
	\end{tabular}
	\caption{(Color online.) Extraction of the first intercepts of $5$ different values of $c$ for 4 values of $B^{(y)}(t_0)/J_0$: (a) $0.94$, (b) $0.74$, (c) $0.49$, and (d) $ 0.35$. The values of the intercepts $c$ are: $-0.0002$ (black circles), $-0.0005$ (red squares), $-0.001$ (blue triangles ), $-0.0015$ (purple plus signs), and $ -0.002$ (green diamonds). In addition, the dashed lines show a power law fit, $|R_0 - R_j| \propto t^{\gamma}|$, for each of these values. The $-0.0002$ power law fits are the most consistent among the other intercepts. Interestingly, the $-0.0005$ intercept is able to recover the power law behavior in (d). Here the $x$-axis is labeled at the spin's relative distance from $i = 0$, $|R_0 - R_j|$.}

	\label{fig:Intercept}
\end{figure*}

We next plot the first intercepts of when $G_{x x , ij}^{\rm ret}(t,0)=c$ for $5$ different values of $c$, as shown in Fig.~\ref{fig:Intercept}. The values we chose are: $c = -0.0002$, $-0.0005$, $-0.001$, $-0.0015$, and $ -0.002$ for the same transverse magnetic fields as previously used.
The $G_{x x , ij}^{\rm ret}(t,0) = -0.0002$ intercept consistently seems to behave closest to a power law, since the black circles do not vary far from the fitted dashed line. However, for the higher intercepts $G_{x x , ij}^{\rm ret}(t,0) = -0.001$, $-0.0015$, and $ -0.002$, the power law fits well only for Fig.~\ref{fig:Intercept}~(a). As the transverse magnetic field decreases, the oscillations begin to dominate the pure state retarded Green's function and the power law fit begins to fail. Although for $G_{x x , ij}^{\rm ret}(t,0) = -0.0005$, a power law fit can be calculated for $B^{(y)}(t_0)/J_0 = 0.94$, $0.74$, and $ 0.35$, as shown in Fig.~\ref{fig:Intercept}~(a, b, and d), it cannot for the intermediate transverse magnetic field in Fig.~\ref{fig:Intercept}~(c). Since the Lieb-Robinson bound is determining the initial response, we expect this approach to work best in the limit as the intercept $c\rightarrow0$. But this limit would be essentially impossible to reach experimentally, where one wants to use as large an intercept as possible.

In the short time limit, we expect the pure state retarded Green's function to be proportional to $J_{ij}$ multiplied by a spin-spin expectation value, as described by the third spectral moment in Eq.~(\ref{eq:moments}c) for $i\neq j$. And following from Eq.~(\ref{eq:approx}), the pure state retarded Green's function is then proportional to $|R_0 - R_j|^{-\alpha}t^3\langle \psi_0|\sigma_0^{(y)}\sigma_j^{(y)}|\psi_0\rangle$. If this spin-spin expectation value were a constant with distance (which would occur for a fully ordered ferromagnetic state), then one could predict the power law to approach $3/\alpha$. Instead, we predict that the power law $|R_0-R_j|\propto t^\gamma$ would have $\gamma$ inversely proportional to $\alpha$. This relation shows that increasing $\alpha$ should reduce $\gamma$, which we see in our data, and was also seen in the experiment which approached these bounds in a different way~\cite{monroe_liebrobinson, roos_liebrobinson}. This relationship between the relative distance of the spins to $t$ examines a generalized Lieb-Robinson bound, which puts an upper bound on how quickly information travels down the chain of spins. Note that the original Lieb-Robinson bound was derived using the maximum value (or maximum eigenvalue) of the retarded Green's function, $|| G_{x x , ij}^{\rm ret}(t ) ||$ ~\cite{lieb_robinson}, and we are considering the expectation value of the retarded Green's function given a pure state $| \psi_0\rangle$ that is not typically a single eigenstate.  So the results are a bit different.

In Table~\ref{tab:fits}, we show the power law fits, $|R_0 - R_j| \propto \bar t^{\gamma}$, for the different features discussed previously for $3$ different Ising interaction power laws: $\alpha = 0.9$, $1.00$, and $1.12$. The power law fits for the first local minimum, first local maximum and the first zero crossing vary with little discernible behavior as the transverse magnetic field is decreased. The $G_{x x , ij}^{\rm ret}(t,0) = -0.0002$ gives consistent values for $\gamma$ for all the Ising interactions we considered, but the $\gamma$s are higher than $\alpha$. More than likely the intercept is still set too high and higher order spectral moments are being observed as well. Although from the $G_{x x , ij}^{\rm ret}(t,0) = -0.0002$ fits, the $\gamma$'s are inversely proportional to the $\alpha$'s. The fits for intercepts where $G_{x x , ij}^{\rm ret}(t,0) =- 0.0005$, $-0.001$, $-0.0015$, and $ -0.002$ are less consistent for low transverse magnetic field. The inconsistency and negative $\gamma$ are due to the oscillatory behavior of the pure state retarded Green's function. 

\begin{table*}
\begin{center}
\begin{tabular}{ | l| l | l | l | l | l | l | l | l | l |}

\hline

$B^{(y)}(t)/J_0$ & $\alpha$ & $0.0002$& $0.0005$& $0.001$& $0.0015$ & $0.002$ & first local minimum & first local Maximum & first zero \\ \hline
\multirow{3}{*}{ 0.94 } & 0.90 & 2.47 & 2.41 & 2.43 & 2.44 & 2.43 &  1.34 & 1.78 & 6.09\\
& 1.00 & 1.91 & 1.95 & 1.95 & 1.95 & 1.96 &  2.12 & 3.211 & 0.892\\
& 1.12 & 1.72 & 1.77 & 1.77 & 1.77 & 1.78 & 2.21 & 2.92 & 1.31 \\
\hline
\multirow{3}{*}{ 0.74 } & 0.90 & 2.39 & 2.38 & 2.33 & 2.28 & 2.25 & 8.94 & --- & 9.81 \\
& 1.00 & 1.93 & 1.94 & 1.93 & 1.93 & 1.92 & 5.26 & --- & 1.19 \\
& 1.12 & 1.75 & 1.76 & 1.76 & 1.75 & 1.75 & 1.21 & 2.05 & 7.05 \\
\hline
\multirow{3}{*}{ 0.49 } & 0.90 & 2.64 & 2.77 & 2.82 & 2.83 &  2.81 & 1.41 & 0.85 & ---\\
& 1.00 & 1.87 & 1.0 & --- & --- & --- &   0.38 & 0.30 & 3.20\\
& 1.12 & 1.69 & 1.59 & 1.01 & --- & --- & 4.29 & --- & 2.88 \\
\hline
\multirow{3}{*}{ 0.35 } & 0.90 & 1.96 & --- & --- & --- & --- & 1.22 & 0.82 & 1.71 \\
& 1.00 & 1.97 & 1.95 & --- & --- & --- & --- & 0.51 & 0.31 \\
& 1.12 & 1.59 & 1.72 & 1.75 & --- & --- & --- & 0.75 & 1.2 \\
\hline
\end{tabular}
\end{center}
\caption{ Power law fits $|R_0 - R_j| \propto \bar t^{\gamma}$ for $3$ different ranges of Ising interaction, $\alpha = 0.90$, $1.00$, and $1.12$. Each are calculated at $B^{(y)}(t_0)/J_0 = 0.94$, $0.74$, $0.49$, and $ 0.35$. The calculated $\gamma$ are for the following features: $G_{x x , ij}^{\rm ret}(\bar t,0) = -0.0002$, $-0.0005$, $-0.001$, $-0.0015$, and $ -0.002$; first local minimum; first local maximum; and the first zero crossing. Here, the ---'s are for $\gamma$'s that inadequately fit the data or are negative. The $G_{x x , ij}^{\rm ret}(\bar t,0) = -0.0002$ case gives a consistent $\gamma$ for the $3$ Ising interaction ranges. Other cases do not work as well. }
\label{tab:fits}
\end{table*}

\subsection{ Energy spectrum }

\begin{figure*}
	\centering
	\begin{tabular}{ c c }
		\includegraphics[scale=0.065]{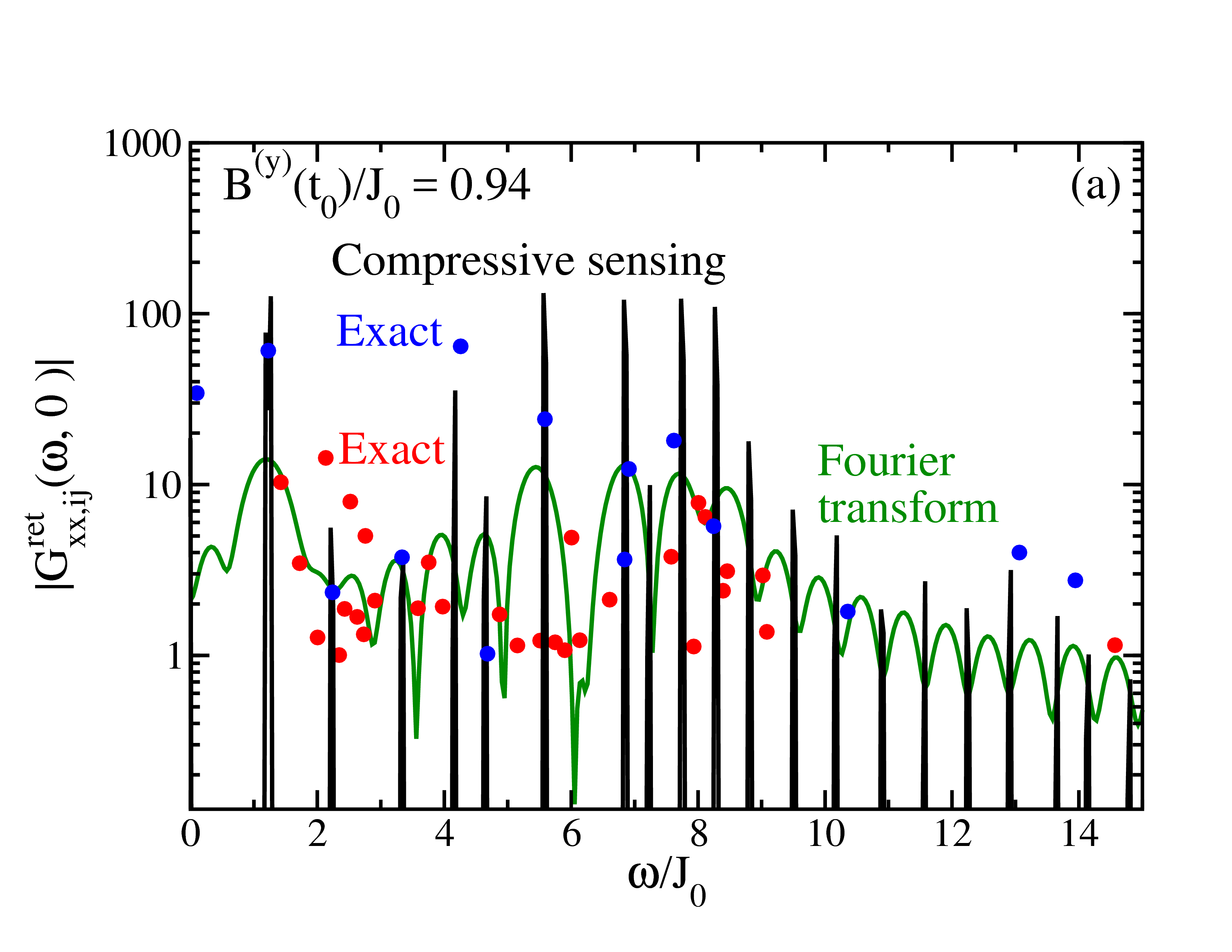} & \includegraphics[scale=0.065]{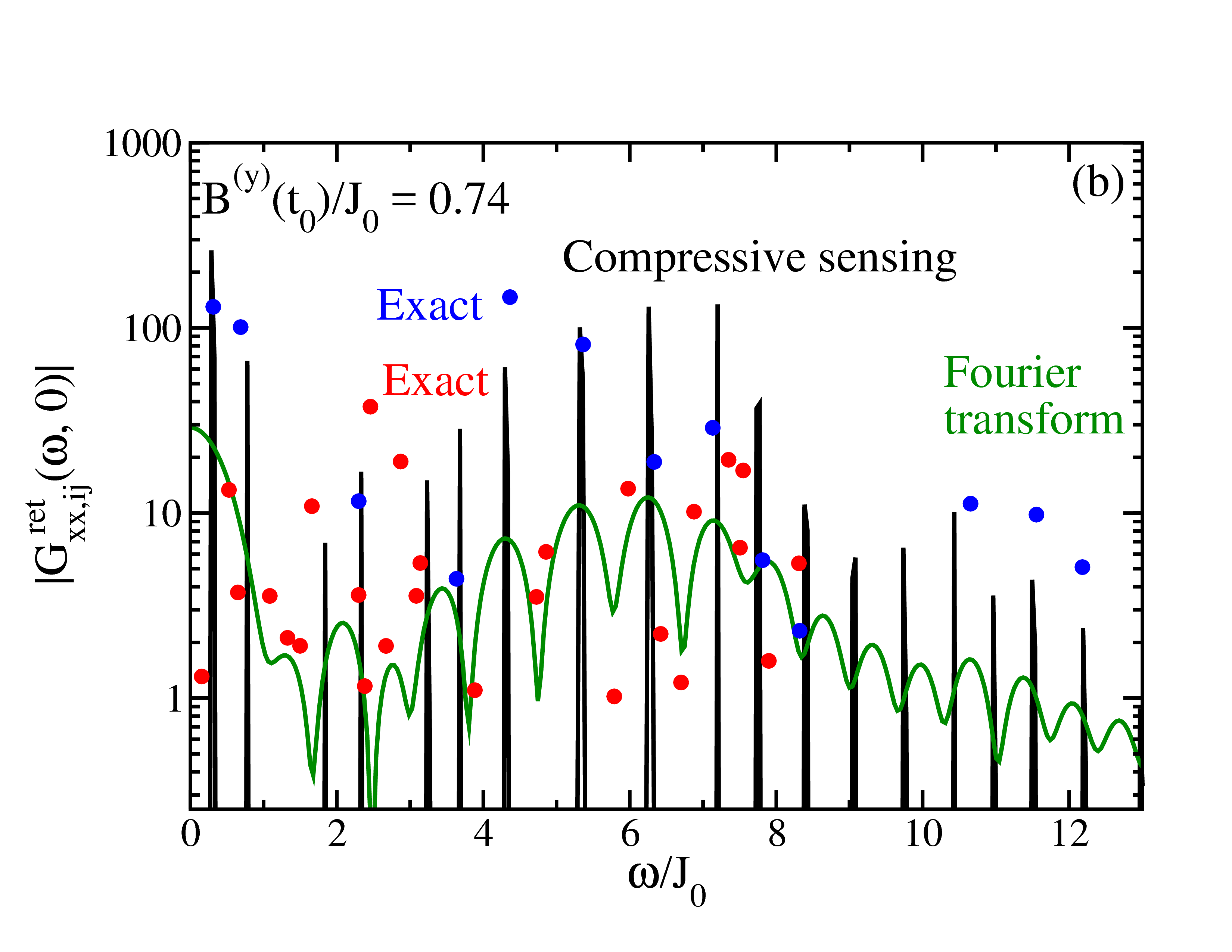}\\
		\includegraphics[scale=0.065]{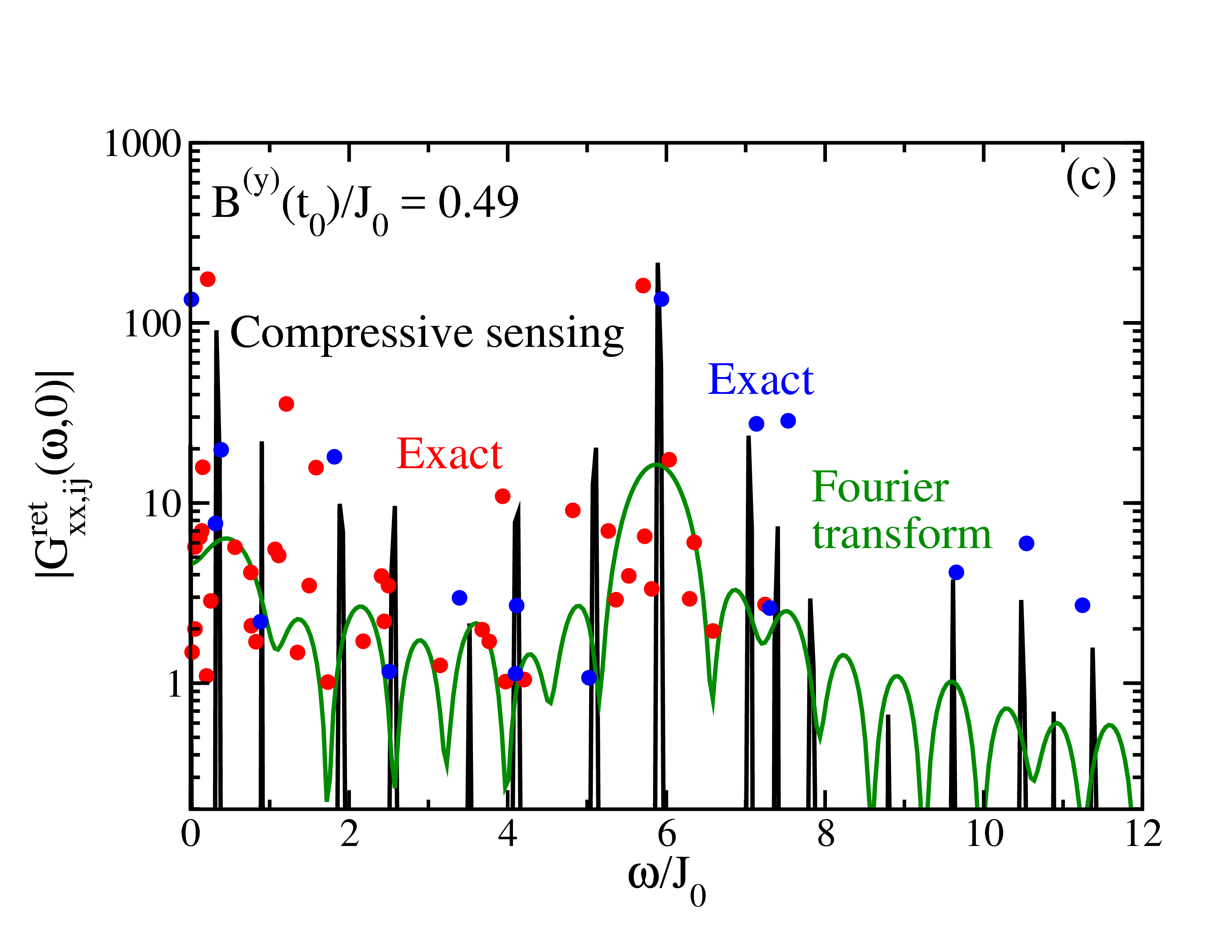} & \includegraphics[scale=0.065]{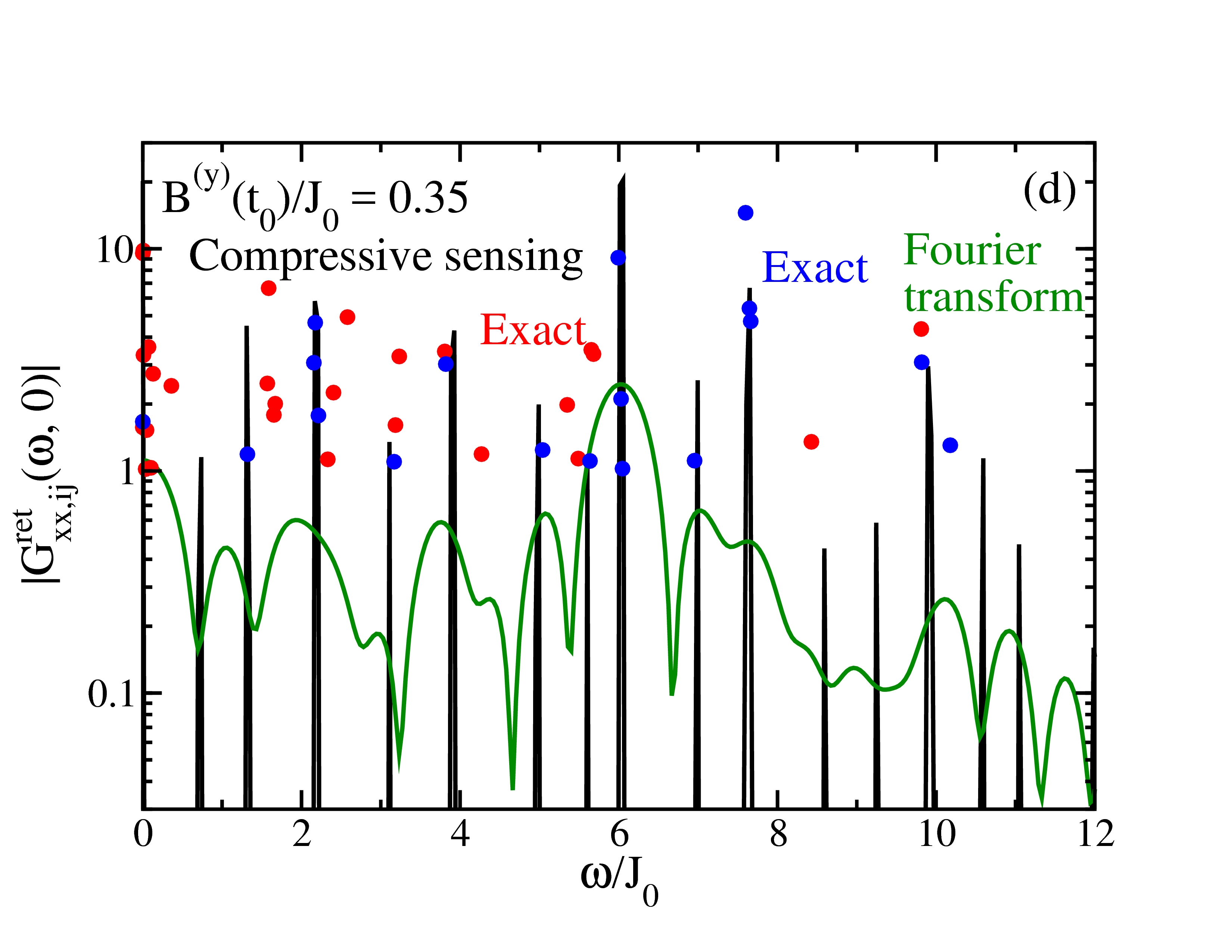}
	\end{tabular}
	\caption{(Color online.) Fourier transform of the $i = j = 0$ pure state retarded Green's function of the data taken in the time interval of $[0,6]$ms. The results of the compressive sensing (black line) are compared to the scaled coefficients of the Lehmann representation,  $1024 G_{x x , 0 0}^{\rm ret}(t,0)$ (blue and red circles), and to the partial Fourier transform (green line). The blue circles are energy differences that are associated with a delta function peak and the red circles are those that cannot be easily associated with a delta function peak. There are also spurious delta functions peaks that are due to the addition of two energy differences. }
	\label{fig:Fourier}

\end{figure*}

We present numerical examples for extracting the energy spectrum via Fourier transformation of the $i = j = 0$ data. The measurement, or calculation, of the energy spectrum has been previously examined experimentally~\cite{monroe_spectroscopy, roos_spectroscopy} and theoretically~\cite{Knap, us_spectroscopy}. The Fourier transform of the $i = j = 0$ data in Fig.~\ref{fig:TimeSlice} is performed over a time interval of $[0,6]$ms for $\alpha = 1.00$ with $B^{(y)}(t_0)/J_0 = 0.94$, $0.74$, $0.49$, and $ 0.35$. As described above, the Fourier transformation is performed by using compressive sensing. We employ $M=64$ time steps that map to $N_{step}=1024$ steps in the frequency domain in Fig.~\ref{fig:Fourier} and compare the resulting delta function peaks to scaled coefficients from the Lehmann representation of $1024 G_{x x , 0 0}^{\rm ret}(t,t_0)$.  
In Fig.~\ref{fig:Fourier}, the blue dots represent scaled coefficients that can be identified to a delta function peak and the red dots are the coefficients that cannot be readily identified with a delta function peak. When the number of steps in the compressive sensing data, $M$, is increased, the number of associated scaled coefficients increases as does the accuracy. Note though that the majority of the red dots do cluster near blue dots and the compressive sensing might not be able to distinguish between the different peaks. There are a few spurious high frequency delta function peaks that are due to frequencies being added together, however identifying what those frequencies are is not easily done. 
So the Green's function is showing that it contains much information about the spectra, unfortunately it is difficult to extract this data from a series of temporally short experimental runs.
Here we used a basic compressive sensing scheme but a more advanced compressive sensing algorithms ({\it i.~e.} basis pursuit) might be able to further reduce the number of measurements as a function of time, reduce the time interval, and produce more accurate delta function peaks. In addition there are compressive sensing algorithms that can reduce the effects of noise and counting statistical error, ({\it i.~e.} basis pursuit denoise). 

\section{Conclusion} 

In this work, we investigated the nonequilibrium behavior of the pure state retarded spin-spin Green's function produced by a variant of a Ramsey spectroscopy protocol by exploring its application to the transverse-field Ising model as simulated in a linear Paul trap. First, we showed that the Lehmann representation of the pure state retarded Green's function is generalized and we determined the first three spectral moment sum rules. We proceeded to present numerical examples of the Ramsey spectroscopy as a function of time. We then extracted the various features to simplify the pure state retarded Green's function behavior. The features we chose to extract are: the first local minimum; the first local maximum; the first zero crossing; and when $G_{x x , ij}^{\rm ret}(t,0) = c$. From these features, we fit power laws to investigate Lieb-Robinson bounds. The feature that gave the most consistent power laws was the smallest intercept, however the resulting power law was higher than expected. This is most likely due to the intercept being too high and the oscillatory behavior affecting the signal. The final example was to Fourier transform the measurement as a function of time into the frequency domain. Compressive sensing was used to extract the excitation energies weighted by the matrix elements of the generalized Lehmann representation. This analysis was not able to extract all of the energy differences. Additionally there are spurious high frequency delta function peaks that are most likely due to two frequencies being added together. 

What are the experimental implications of these results? It appears that it would be difficult to use this method to extract generalized Lieb-Robinson bounds, because the results work best for small intercepts, but experimental error would make the data there very noisy. Similarly, one can extract some of the excitation energy differences, but not all of them because there are too many of them, hence it is not clear precisely what one would do with the experimentally measured subset of data. Perhaps, the most interesting aspect of this Green's function is the Green's function itself. After all, it is surprising to be able to extract a retarded Green's function from an experimental measurement, and a full knowledge of the Green's function allows for a wealth of different information to be determined about the system.
Indeed, this is likely the most important result of the Ramsey experiment in these trapped ion systems. This becomes even more interesting if one examines the more nonequilibrium case where the Hamiltonian continues to change between times $t_0$ and $t$. Unfortunately, it isn't clear precisely what one would use that data for.  

Another interesting question is the following: In cases where the pure state $|\psi_0\rangle$ represents a thermal distribution well, in the sense that the coefficients $|C_n|^2$ are nearly proportional to the Boltzmann factor~\cite{lim}, then does the pure state retarded spin-spin Green's function represent the thermally
averaged retarded Green's function well? One might expect this to be true, because the diagonal elements in the summation will closely resemble the trace employed in the calculation of the thermal Green's functions, and the off diagonal elements should become small as the system size becomes large due to cancellations from the complex phases.

We hope future studies will clarify these issues.
 
\section*{Acknowledgments}
J. K. F. and B. Y. acknowledge support from the National Science Foundation under grant number PHY-1314295. J. K. F. also acknowledges support from the McDevitt bequest at Georgetown University. B. Y. also acknowledges support from the Achievement Rewards for College Students Foundation.

\end{document}